\documentclass[12pt]{article}
\usepackage{amsfonts,amsmath,amssymb,bm}
\usepackage{graphicx}
\usepackage{float}
\usepackage{hyperref}
\usepackage[paper=letterpaper,margin=.85in]{geometry}
\usepackage{xcolor}
\usepackage[mathscr]{euscript}
\usepackage{physics}
\usepackage{tikz}
\usepackage{mathtools}
\usepackage{slashed}
\usetikzlibrary{arrows.meta,decorations.pathmorphing,decorations.markings,patterns,patterns.meta}
\tikzset{
midarr/.style={postaction={decorate},decoration={markings,mark=at position 0.55 with {\arrow{>}}}},
midcirc/.style={
    postaction={decorate},
    decoration={markings, mark=at position 0.55 with {\draw[fill=gray] circle (0.15);}}
  },
bprop/.style={thick,dashed,midarr},
yline/.style={thick,midarr},
ylineR/.style={thick,midarr},
ydouble/.style={thick,double,double distance=1.4pt,midarr},
Xdouble/.style={thick,double,double distance=1.4pt,midcirc}
}

\usepackage{comment}
\usepackage[numbers,sort&compress]{natbib}
\graphicspath{{too_many_diagrams/}}

\renewcommand{\tilde}{\widetilde}
\renewcommand{\d}{\mathrm{d}}
\newcommand{\D}{\mathcal{D}}
\newcommand{\pf}{\mathrm{Pf}}
\newcommand{\Pf}{\operatorname{Pf}}
\newcommand{\ii}{\mathrm{i}}
\renewcommand{\dd}{\mathrm{d}}
\newcommand{\ad}{\mathrm{ad}}
\newcommand{\cten}{c_{10}}
\newcommand{\bfss}{\text{BFSS}}
\newcommand{\lr}[1]{\langle #1 \rangle}
\newcommand{\mV}{\mathcal{V}}

\numberwithin{equation}{section}
\def\tr{\operatorname{Tr}}
\def\Tr{\operatorname{Tr}}
\def\qu{\text{quench}}

\numberwithin{equation}{section}

\def\tr{\text{Tr}}
\def\Tr{\text{Tr}}

\begin{document}
\thispagestyle{empty}

\vspace*{2.5cm}
\begin{center}

     {\LARGE \bf An effective field theory approach to the sign problem in BFSS }

\vspace{0.5in}

{\bf Gauri Batra$^1$, Henry W. Lin$^{2,1}$ \& Haifeng Tang$^1$}

\vspace{0.3in}

$^1$Leinweber Institute for Theoretical Physics, Stanford University, Stanford, CA 94305, USA

\vspace{0.1in}

$^2$Joseph Henry Laboratories, Princeton University, Princeton, NJ 08544, USA
                
    \vspace{0.5in}

    \vspace{0.5in}
    
\end{center}

\vspace{0.5in}

\begin{abstract}
The sign problem is a notorious obstacle for classically simulating quantum theories with fermions. We propose an effective field theory method for analyzing the sign problem. 
At high temperatures, a $d$+1 dimensional field theory reduces to a bosonic $d$-dimensional theory; the phase of the Pfaffian in the higher dimensional theory is encoded in an operator in the lower dimensional theory. 
We apply this framework to the D0-brane/BFSS matrix quantum mechanics, where the phase becomes an operator in a bosonic multi-matrix integral. 
Our results show that the continuum theory has a sign problem that persists in the large-$N$ 't Hooft regime. 
However, detecting the sign problem involves going to 10-loop order in the high-temperature expansion. 
This delayed onset follows from the fact that the Pfaffian phase transforms as an $O(9)$ pseudoscalar. 
Furthermore, the relevant diagrams give a numerically small prefactor. Consequently, ignoring the sign problem leads to a relatively small fractional error in thermodynamic quantities for temperatures $T \gtrsim \lambda^{1/3}$. However, at stronger coupling in the 't Hooft regime, the sign problem may become more severe. Finally, we initiate the study of the sign problem in higher-dimensional maximally supersymmetric Yang-Mills theories.
\end{abstract}

\vspace{1in}

\pagebreak

\setcounter{tocdepth}{3}

\tableofcontents

\section{Introduction}

Simulating quantum systems on classical computers is difficult in general.  In Euclidean signature, one can discretize the path integral and try to evaluate the integral by Monte Carlo. 
If the system includes fermions, one can do the Grassmann integrals first, leaving behind an integral over bosonic fields. In practice, the fermions often enter quadratically, so one obtains a Pfaffian which depends on the bosonic fields $X$: %
\begin{align}
    Z = \int D X \, D \psi \, e^{-S(X,\psi)} = \int D X \, e^{-S_\text{bos}(X)} \Pf(M(X)).
\end{align}
However, even in Euclidean signature, the resulting effective ``measure'' may not be positive.\footnote{Even without fermions, the bosonic measure may not be positive, e.g., a gauge theory with a $\theta$ angle or a Chern-Simons term.}
This is the {\it fermion sign problem} (see \cite{Gattringer:2016kco,Nagata:2021ugx} for reviews).  In practice, a workaround is the ``phase-quenched" approximation: one replaces the Pfaffian $\Pf(\mathcal{M})$ by its absolute value $|\Pf(\mathcal{M})|$ to obtain a non-negative measure:
\begin{align}
    {Z_\qu} = \int D X \, e^{-S_\text{bos}(X)} |\Pf(M(X))|.
\end{align}
Sampling from this measure is presumably only reliable when the phase of the Pfaffian does not fluctuate ``too much'' between the dominant configurations in the path integral. If the sign or phase of the Pfaffian fluctuates wildly between different bosonic configurations, it will be difficult to measure any observable accurately, even if one reweights by the phase.  
The goal of this paper is to better quantify the sign problem and to clarify in what situations it might be neglected. We will focus mainly on the D0-brane quantum mechanics \cite{deWit:1988wri} or Banks-Fischler-Shenker-Susskind (BFSS) matrix theory \cite{Banks:1996vh}, although our method of studying the sign problem can be applied to other theories (including higher dimensional super Yang Mills theories).

Despite the potential sign problem, Monte Carlo simulations of the D0-brane/BFSS quantum mechanics have seen remarkable progress \cite{Hanada:2007ti,Catterall:2007fp,Anagnostopoulos:2007fw,Catterall:2008yz, Hanada:2008ez, Catterall:2009xn, Hanada:2011fq,Filev:2015hia,Kadoh:2015mka,Berkowitz:2016jlq,Pateloudis:2022ijr} (see \cite{Taylor:2001vb, Lin:2025iir, YinStringbook} for reviews of BFSS). 
Early simulations found the effects associated with the Pfaffian phase to be small
in the accessible finite-$N$, finite-cutoff regimes
\cite{Anagnostopoulos:2007fw,Catterall:2008yz,Catterall:2009xn,
Hanada:2011fq,Filev:2015hia}. Although nonzero phase fluctuations were observed in some ensembles, they were either small or had little effect on the simple
observables studied\footnote{See Appendix C of \cite{Hanada:2011fq} for a study of the sign problem for the SU(3) BFSS model; there, they presented evidence that phase re-weighting does not affect observables such as moments of $\tr X^I X^I$ even though the phase has significant fluctuations.}. For example, \cite{Filev:2015hia} reported
$\langle\cos\theta_{\mathrm{Pf}}\rangle\approx1$ at $N=3$ and finite lattice spacings across a range of temperatures. They conjectured that the deviation from 1 was a lattice artifact, e.g., $\langle \cos \theta_{\mathrm{Pf}} \rangle = 1 $ in the continuum limit. These results motivated subsequent Monte Carlo work on BFSS to employ the phase-quenched approximation without phase reweighting.

Interestingly, the results with the phase-quenched approximation agree with the holographic predictions for the black hole thermodynamics (within numerical uncertainties) and lead to interesting predictions for $\alpha'$ corrections. This suggests that the sign problem is not severe, at least for thermodynamic observables, but one could wonder whether it is parametrically under control.

In this paper, we study the fermion sign problem in the BFSS matrix model analytically, using the high-temperature expansion. At high temperatures, the path integral is dominated by the bosonic zero modes. This 0D integral is a bosonic version of the IKKT model \cite{Ishibashi:1996xs, Kawahara:2007ib, Gur-Ari:2015rcq}; it is bosonic as the anti-periodic boundary conditions for the fermions mean that the fermionic modes have non-zero Matsubara frequencies. Thus at infinite temperature $\lambda^{1/3} \beta \to 0$ holding $N$ fixed, there is no sign problem. We can therefore hope to study the sign problem in perturbation theory in the coupling $\lambda \beta^3 = (g^2_\text{YM} N) \beta^3$. 

We find that the sign problem only appears at a relatively high order in the perturbative expansion. Indeed the leading effect is a 10-loop diagram; see Figures \ref{fig:planar_annulus} and \ref{fig:replica_nonperturbative_internal}. In other words, we will show that when diagnosing the sign problem via $\ev{\cos \theta_\text{Pf}}$, all diagrams involving fewer loops will automatically cancel. Although initially surprising, this cancellation is explained by symmetry principles. While the finite temperature theory only has an SO(9) symmetry, the zero-mode matrix integral has an {\it enhanced} $O(10)$ symmetry at leading order in the high temperature expansion. An interesting subgroup of $O(10)$ is $O(9) \times \mathbb{Z}_2$ where the $\mathbb{Z}_2$ reflects the gauge field $A_0$. As we will see, $\theta_{\Pf}$ is an $O(9)$ pseudoscalar and is $\mathbb{Z}_2$-odd. In terms of the dimensionless variables $Y_\mu$ of the bosonic zero-mode integral, the leading single-trace operator consistent with these symmetries is  $\mathcal{T}_{10} = \epsilon^{i_1 \cdots i_9} \Tr( Y_0 Y_{i_1} \cdots Y_{i_9})$. 
The fact that this operator involves 10 insertions of $Y$ explains why one has to go to high orders in perturbation theory.

In particular, for large $N$, we show that $\log \ev{\cos\theta} \propto \ev{\mathcal{T}_{10}^2}$, yielding
\begin{align} \label{eqIntro1}
    \ev{\cos \theta_{\pf}}  &= \frac{Z_\bfss}{Z_\qu} \approx  \exp \left[-  N^2 f(\lambda^{1/3} \beta) \right], \quad
    f = c \lambda^{5} \beta^{15} + O(\beta^{33/2}) , \quad c = 9.1 \times 10^{-9}.
\end{align}
We will see that the numerically small value $c\sim 10^{-8}$ also stems from the fact that 10 is a large number.

This falsifies the conjecture \cite{Filev:2015hia} mentioned above that $\ev{\cos \theta} = 1$ in the continuum theory. Indeed, in the 't Hooft regime $N \to \infty$ holding fixed $\lambda \beta^3$, our results imply that $\ev{\cos \theta} \to 0$. Nevertheless, for temperatures $T \gtrsim 1$ in the 't Hooft regime, there is a sense in which the sign problem is mild: {\it it leads to a small fractional error in the thermodynamics}.
This is due to the small numerical value of $c$ and to the high power of $\beta$ that appears. To illustrate our results, let us consider a typical value of the temperature for Monte Carlo simulations $T/\lambda^{1/3} = 0.35 $, we find a fractional error in the free energy of about 10\%, see \eqref{sugra_predict} and the discussion below it. 
Of course, as we lower the temperature (to make contact with the strong coupling/gravity regime), the sign problem seems to become more severe, although one would need to compute higher order corrections in high-temperature perturbation theory (and ideally resum the series) in order to say anything definitive.

Our treatment may be viewed as a new application of the standard high-temperature effective action approach to thermal field theories, which has been studied for gauge theories and CFTs in, e.g., \cite{Appelquist:1981vg,Braaten:1995cm,Kajantie:1995dw, Banerjee:2012iz,Jensen:2012jh, Benjamin:2023qsc}. 

This paper is organized as follows. In section~\ref{sec:pert} we set up the high-temperature expansion of BFSS, derive the leading Pfaffian phase operator, and give the symmetry argument for why the operator first appears at tenth order in the perturbative expansion. In section~\ref{sec:largeN} we analyze the large $N$ limit of this operator and estimate its coefficient in the large $N$, large $D$ expansion. In section~\ref{sec:replica} we rederive the same result using a replica formulation of the phase-quenched theory. In section~\ref{sign_higherD} we generalize our results to higher dimensional SYM theories. 
In section~\ref{sec:discussion} we discuss the implications for Monte Carlo simulations and some future directions. The paper is complemented by several appendices which discuss more technical details.

\section{Setup and symmetry argument}
\label{sec:pert}
Our main interest in this work is the gauged BFSS model at temperature $T = 1/\beta$, which is defined by
the Euclidean action 
\begin{equation}
\label{eq:BFSS}
S = \frac{N}{\lambda} \int_0^\beta \dd t\; \tr\left[
  \frac12 (D_t X_i)^2
  + \frac12 \psi_\alpha D_t \psi_\alpha
  - \frac14 [X_i, X_j]^2
  - \frac12 \psi_\alpha (\gamma_i)_{\alpha\beta} [X_i, \psi_\beta]
\right].
\end{equation}
Here $\lambda = g^2 N$ is the 't Hooft coupling, $D_t = \partial_t - \ii [A(t), \cdot\,]$ is the covariant derivative,
$A(t)$, $X_i(t)$ ($i=1,\ldots,9$), and $\psi_\alpha(t)$
($\alpha = 1,\ldots,16$) are $N\times N$ Hermitian matrices,
and $\gamma_i$ are $16\times 16$ real symmetric matrices obeying
$\{\gamma_i,\gamma_j\} = 2\delta_{ij}$.
The bosons are periodic and the fermions are antiperiodic on the thermal circle $[0,\beta]$. At finite temperature the supersymmetries are broken but there is still an unbroken $SO(9)$ $R$-symmetry.

Our Fourier conventions are
\begin{align}
X_i(t) &= \sum_{n\in\mathbb{Z}} X_{i,n}\,e^{\ii\omega_n t},
\qquad
\omega_n=\frac{2\pi n}{\beta},
\\
\psi_\alpha(t) &= \sum_{r\in\mathbb{Z}+\frac12}\psi_{\alpha,r}\,e^{\ii\omega_r t},
\qquad \omega_r=\frac{2\pi r}{\beta},
\end{align}
where $n\in\mathbb Z$ labels bosonic Matsubara modes and
$r\in\mathbb Z+\tfrac12$ labels fermionic Matsubara modes.
We will use the following conventions for zero modes throughout this note:
\begin{align}
X_i(t) &= X_{i,0} + (\text{non-zero Matsubara modes}),\\
A(t) &= A_0 \quad \text{(static gauge)}.
\end{align}
To go to static gauge, we should carry out the Faddeev-Popov procedure. This leads to a measure for $A_0$ that is derived in Appendix~\ref{app:FP}, and is the Haar measure for the unitary matrix that implements the gauging of $SU(N)$. Since it is positive it will not play an important role in our discussion of the phase\footnote{However, it is important that at high temperatures we get a standard Vandermonde interaction for $A_0$ so that the resulting zero-mode model is $O(10)$ invariant.}.

At high temperatures, all the Matsubara non-zero modes become very massive. Therefore we can integrate them out; the resulting theory is a path integral over the bosonic zero modes. 
At tree-level, this just reproduces the classical statistical mechanics limit of the model, which is sometimes referred to as the bosonic IKKT model; below we simply refer to it as the zero-mode theory. This theory has the action
\begin{align}
  S_{0m}
  &=  \frac{N \beta}{\lambda} \,\tr\left(
    -\frac14 [X_{i,0},X_{j,0}]^2
    -\frac12 [A_0,X_{i,0}]^2
  \right)
  \nonumber\\
  &= -\frac{N}{4}\,\tr\left([Y_\mu,Y_\nu]^2\right),
  \qquad Y_\mu=(Y_0,Y_i),
  \label{eq:bIKKT}
\end{align}
where Greek indices run over $\mu=0,\ldots,9$, and we have introduced the
dimensionless zero modes
\begin{equation}
\label{eq:Ydef}
Y_0=\left(\frac{\beta}{\lambda}\right)^{1/4}A_0,\qquad
Y_i=\left(\frac{\beta}{\lambda}\right)^{1/4}X_{i,0}.
\end{equation}
From this action we see that this theory has an enhanced symmetry at high temperatures; it is the $O(10)$ symmetry under which $Y_\mu$ transforms as a vector. Now, the idea is that at leading order in the high temperature expansion, we may replace
\begin{align}
    Z_\text{BFSS} = \int \D A\,\D X\;\Pf(M)\,e^{-S}
    \approx \int \d Y_\mu\; e^{-S_{0m}(Y)} \mathcal{P}(Y;\beta),
\end{align}
 up to a $\beta$-dependent Jacobian that is independent of the matrix configuration. Here $\mathcal P(Y;\beta)$ denotes the fermion Pfaffian evaluated on the zero-mode background.

To compute this Pfaffian, it is useful to write
the fermion action in Matsubara space as
\begin{align}
S_f &= \frac{N}{2\lambda}\sum_{r,s} \psi^a_{\alpha,-r}\,
\mathcal{M}_{r\alpha a,\;s\beta b}\,\psi^b_{\beta,s},\\
\label{eq:Mstatic}
\mathcal M_{r\alpha a,\,s\beta b}
&= \big(2\pi\ii r\,\delta_{ab} - \ii\beta(\ad \, A_0)_{ab}\big)\delta_{rs}\delta_{\alpha\beta}
- \beta(\ad \, X_{i,r-s})_{ab}(\gamma_i)_{\alpha\beta},
\end{align}
Here we expanded \(\psi=\psi^aT_a\), and the adjoint basis is normalized as
\begin{align}
    \tr(T_aT_b)=\delta_{ab},
    \quad
    [T_a,T_b]=\ii f_{abc}T_c .
\end{align}
We also defined
\begin{align}
    (\ad \, Y)_{ab}=\ii f_{acb}Y^c,
    \quad Y=Y^cT_c.
\end{align}
The overall factor of
$N/\lambda$ in $S_f$ has not been included in $\mathcal M$; it changes
$\Pf\mathcal M$ by a background-independent power of $N/\lambda$ and therefore
does not affect the phase. 
As explained before, we have written the action in static gauge $A(t)=A_0$. %
The associated
Faddeev-Popov factor
$\Delta_{\rm FP}(\alpha)=\prod_{i<j}\sin^2\!\big((\alpha_i-\alpha_j)/2\big)$
(derived in Appendix~\ref{app:FP}) is real and non-negative and does not contribute to
the phase of the bosonic measure, so the only possible complex phase comes
from the fermion Pfaffian.

The Pfaffian phase is then given by
\begin{equation}
\label{eq:logPfTrlog}
\theta = \Im \log\Pf\,\mathcal M \;=\; \Im \tfrac12\,\Tr\log\mathcal M,
\end{equation}
where the trace runs over the combined Matsubara, adjoint, and spinor
indices. 

\subsection{Symmetry analysis of the Pfaffian phase}%
\label{sec:reflection}
{In this section, by analyzing some properties of the above matrix $\mathcal M$, we show how $\theta$ transforms under an $O(9)\times \mathbb{Z}_2$ subgroup of the enhanced $O(10)$ symmetry of the zero-mode integral.}

First note that 
\begin{equation}
    (A_0^c)^*=A_0^c,
    \qquad
    (X_{i,n}^c)^*=X_{i,-n}^c.
\end{equation} 
Since the structure constants are real, this implies
\begin{equation}
\label{eq:adstar}
(\ad \, A_0)^*=-\ad \, A_0,\qquad (\ad \, X_{i,n})^*=-\ad \, X_{i,-n}.
\end{equation}
Now, consider the following two $\mathbb Z_2$ operations on the bosonic background:
\begin{align}
\label{eq:R0def}
\mathcal T: \quad & A_0\mapsto -A_0,\qquad X_i(t)\mapsto X_i(-t),\\
\label{eq:Rdef}
R: \quad &  A_0\mapsto A_0,\qquad X_i(t)\mapsto -X_i(t).
\end{align}
These transformations act on Fourier modes as $R:\,X_{i,n}\mapsto -X_{i,n}$ and
$\mathcal T:\,A_0\mapsto -A_0$, $X_{i,n}\mapsto X_{i,-n}$.  Here \(R\) is an improper \(O(9)\) rotation, while \(\mathcal T\) is the
Euclidean time reflection acting on the spatial matrices. As we will explain, these
transformations take the Pfaffian to its complex conjugate:
\begin{align}
\label{eq:PfR0full}
\Pf\,\mathcal M(\mathcal T(A_0),\,\mathcal T(X)) &= \big(\Pf\,\mathcal M(A_0,X)\big)^*,\\
\label{eq:PfRfull}
\Pf\,\mathcal M(A_0,R(X)) &= \big(\Pf\,\mathcal M(A_0,X)\big)^*.
\end{align}
The first identity is most transparent.  Using \eqref{eq:adstar} and
conjugating each piece of \eqref{eq:Mstatic}, we get
\begin{equation}
\mathcal M(A_0,X)^*_{rs}
= -\big(2\pi\ii r\,\delta_{ab} + \ii\beta(\ad \, A_0)_{ab}\big)\delta_{rs}\delta_{\alpha\beta}
+ \beta(\ad \, X_{i,-(r-s)})_{ab}(\gamma_i)_{\alpha\beta}.
\label{eq:conjM}
\end{equation}
On the other hand, acting with $\mathcal T$ on \eqref{eq:Mstatic} gives
\begin{equation}
\label{eq:MR0matrix}
\mathcal M(\mathcal T\,A_0,\,\mathcal T\,X) \;=\; -\mathcal M(A_0,X)^*.
\end{equation}
The Pfaffian of an antisymmetric matrix of even size $D$ obeys
$\Pf(-M) = (-1)^{D/2}\Pf(M)$, and with $D/2$ even here\footnote{The per-Matsubara-pair
dimension $16(N^2-1)$ is even.}, \eqref{eq:PfR0full} follows.

To see the $R$ transformation \eqref{eq:PfRfull}, let us again consider \eqref{eq:conjM}.
Consider a relabelling of the matrix indices
$r\to -r$, $s\to -s$, which restores the
kinetic term $2\pi\ii r\to 2\pi\ii(-r)$ and reverses the Matsubara
frequency of the $X$ field $-(r-s)\to r-s$.
This relabeling can be implemented by conjugating $\mathcal{M}$ by the matrix $P_{rs}\equiv\delta_{r,-s}$.
The result is
\begin{align}
(P\,\mathcal M(A_0,X)^*\,P)_{rs} \;&=\; 
 \big(2\pi\ii r\,\delta_{ab} - \ii\beta(\ad \, A_0)_{ab}\big)\delta_{rs}\delta_{\alpha\beta}
+ \beta(\ad \, X_{i,r-s})_{ab}(\gamma_i)_{\alpha\beta},\\
&=
\mathcal M(A_0,-X).
\end{align}
Since $\Pf(PM^*P)=\Pf(M^*)=(\Pf M)^*$\footnote{The matrix $P$ exchanges each fermionic Matsubara mode $r$ with the mode $-r$. Since $r\in\mathbb Z+\tfrac12$, no mode is fixed by this
exchange. For each pair $(r,-r)$, $P$ acts on the internal adjoint--spinor
space as
\begin{equation}
    P_{(r,-r)}
    =
    \begin{pmatrix}
        0 & \mathbf 1_{d_{\rm int}}\\
        \mathbf 1_{d_{\rm int}} & 0
    \end{pmatrix},
    \qquad
    d_{\rm int}=16(N^2-1).
\end{equation}
This block has determinant $(-1)^{d_{\rm int}}$. Since $d_{\rm int}$ is even,
each Matsubara pair contributes $+1$ to the determinant, implying $\det P=1$.},
\eqref{eq:PfRfull} follows.
We conclude that
$\theta=\Im\log\Pf\,\mathcal M$ is an $O(9)$ pseudoscalar and odd under $\mathcal T$. 

\subsection{High temperature expansion}
\label{sec:hightemp}
We will now identify the leading operator in high-temperature perturbation theory, with the selection rules implied by the symmetry analysis performed in the previous subsection.
At high temperature the bosonic non-zero modes are parametrically heavier
than the zero modes, and we may integrate out these ``fast modes.'' At leading order in the high temperature expansion, this amounts to evaluating 
$\Pf\,\mathcal M$ on a static background.  Setting $A_{n\neq 0}=0$,
$X_{i,n\neq 0}=0$, the kernel \eqref{eq:Mstatic} becomes block-diagonal
in Matsubara frequency:
\begin{equation}
\label{eq:M0}
\mathcal{M}^{(0)}_{rs}
= \delta_{rs}
\big(2\pi\ii r\,\mathbf{1}   - \beta K 
\big), \qquad K = \ii\,\ad \, A_0\otimes\mathbf{1}_{16}
  + \ad \, X_{i,0}\otimes\gamma_i.
\end{equation}
On this static background, the
operations $R$ and $\mathcal T$ of section~\ref{sec:reflection} reduce to
$(Y_0,Y_i)\mapsto(Y_0,-Y_i)$ and $(Y_0,Y_i)\mapsto(-Y_0,Y_i)$ respectively
(the time-reversal in $\mathcal T$ acts trivially on constants); these two improper elements of $O(10)$, together with the
$SO(9)$ $R$-symmetry, generate $O(9)\times\mathbb Z_2\subset O(10)$.

The Pfaffian phase on the static background is
\def\im{\mathop{\textrm{im}}}
\begin{align}
 \theta  &= 
\frac12 \im \sum_{r \in \mathbb{Z} + 1/2} \Tr_{\ad,\gamma}
\log\!\left(1 - \frac{\beta }{2\pi\ii r}K\right)  %
=
-\im \sum_{r \in \mathbb{Z} + 1/2} \sum_{n=1}^\infty  \frac{1}{2n} \Tr_{\ad,\gamma}
\left( \frac{\beta }{2\pi\ii r}K\right)^n, \label{eq:TrlogExpand}
\end{align}
where we have dropped a constant independent of $\beta$. Now, the results of section~\ref{sec:reflection} imply that in the above expansion in powers of $K$, the first non-zero term that contributes to $\theta$ is of order $\sim K^{10}$, 
since we need nine gamma matrices to make an $O(9)$ pseudoscalar\footnote{This follows from $\Tr_\gamma(\gamma_{i_1}\cdots\gamma_{i_9})=16 \epsilon_{i_1\cdots i_9}$.} and we need at least one factor of $A_0$ to make the operator $\mathbb{Z}_2$ odd. Thus, the leading operator is %
\begin{equation}
\theta = \im\left\{ -\frac{\beta^{10}}{20}
\sum_r \frac{1}{(2\pi\ii r)^{10}}
\sum_{k=0}^{9}
\Tr_{\ad,\gamma}\!\left[
\big((\ad \, X_{j,0})\gamma_j\big)^k(\ii \, \ad \, A_0)
\big((\ad \, X_{j,0})\gamma_j\big)^{9-k}
\right] \right\} .
\end{equation}
By cyclicity of the combined (adjoint $\otimes$ spinor) trace, the sum over $k$ just produces a factor of 10, so we are left with:
\begin{equation}
\theta
=  \im\Bigg\{ \frac{\beta^{10}}{2}\,\ii \frac{31}{1451520}
\underbrace{\Tr_\gamma(\gamma_{i_1}\cdots\gamma_{i_9})}_{
16\,\epsilon_{i_1\cdots i_9}}\;
\tr_{\ad}\!\left((\ad \, A_0)(\ad \, X_{i_1,0})\cdots(\ad \, X_{i_9,0})\right)\Bigg\}.
\end{equation}
Passing to the dimensionless zero modes \eqref{eq:Ydef}, and defining
\begin{align}
\label{eq:O10def}
\mathcal{O}_{10}
&= \epsilon_{i_1\cdots i_9}\,
\tr_{\ad}\!\left((\ad \, Y_0)(\ad \, Y_{i_1})\cdots(\ad \, Y_{i_9})\right),
\end{align}
the dimensionful trace in the previous line is therefore
$(\lambda/\beta)^{10/4}\mathcal O_{10}$.
The $Y_\mu$ variables have expectation values that are temperature-independent in the high temperature limit.
In terms of these variables, we have
\begin{align} \label{eq:final}
\theta = \cten \,  (\lambda \beta^3)^{5/2} \,\mathcal{O}_{10},
\qquad
\cten=\frac{31}{181440}.
\end{align}
The same selection rules also organize the subleading terms, see Appendix \ref{sec:subleading_T}. We estimate that the next-to-leading term gives a correction that involves 12 $Y$'s and scales like $\theta \sim(\lambda\beta^3)^3 Y^{12}$.

Let us conclude this section by justifying the first equality in \eqref{eqIntro1}. The $O(10)$ symmetry of the zero-mode theory implies that
$\ev{\sin \theta} = 0$, since 
$\theta$ is $\mathbb{Z}_2$ odd under either $R$ or $\mathcal{T}$, whereas the zero-mode measure is even. %
Thus,
\begin{align} \label{eqCosTheta}
    \ev{e^{\ii \theta}}  = \frac{\int D X \, e^{-S_\text{bos}(X)} \Pf(M(X))}{\int D X \, e^{-S_\text{bos}(X)} |\Pf(M(X))|} = \frac{Z_\bfss}{Z_\qu}  = \ev{\cos \theta}.
\end{align}

\subsection{The ungauged model}
For the ungauged BFSS \cite{Maldacena:2018vsr, Berkowitz:2018qhn}, one simply sets the gauge field $A_0 \to 0$ instead of integrating over the gauge field. Equivalently, we can think of $A_0$ as a background gauge field. Our above analysis implies that the Pfaffian phase is an odd function of $A_0$. Thus, 
\begin{align}
A_0 \to 0, \quad \theta = - \theta \mod 2\pi.
\end{align}
We therefore conclude that for every bosonic field configuration $X(t)$, either $\theta = 0$ or $\theta = \pi$ for the ungauged model. This implies that the Pfaffian is real\footnote{See \cite{Nishimura:2000ds} for a related discussion of the Pfaffian in the IKKT model.}  (either positive or negative), unlike in the gauged case where it is complex. 

Since the phase cannot fluctuate continuously, our perturbative approach will fail to detect the sign problem in the ungauged model. Nevertheless, we found static bosonic configurations where the Pfaffian is negative. This seems to imply that the sign problem in the ungauged model is significantly milder than in the gauged model; we leave a more detailed study to future work.
This is particularly interesting as it has been argued \cite{Maldacena:2018vsr, Berkowitz:2018qhn} that the thermodynamics of the gauged vs ungauged model are the same at very strong 't Hooft coupling.

\section{The large \texorpdfstring{$N$}{N} limit}
\label{sec:largeN}
The finite-$N$ result~\eqref{eq:final} implies that one needs to evaluate the moments of the operator $\mathcal{O}_{10}$ to determine $\ev{\cos \theta}$.
In the large-$N$ limit, two simplifications occur:
first, the adjoint trace reduces to a single trace in the fundamental representation at
leading order in~$N$, and
second, the distribution of $\mathcal{O}_{10}$ becomes Gaussian,
so  
\begin{align}
  \ev{e^{\ii \theta}}  = \ev{\cos \theta} \approx \exp \left(-\frac12 \ev{ \theta^2} \right).
\end{align}
We demonstrate these facts below.

\subsection{Simplification of the adjoint trace}

To convert between the adjoint trace and the usual trace, it is convenient to use the Choi--Jamio{\l}kowski isomorphism. This says that we view linear operators as vectors in the doubled Hilbert space: 
$
|X\rangle\!\rangle \equiv \sum_{a,b=1}^N X_{ab}\,|a\rangle\otimes|\bar b\rangle
\in \mathcal H\otimes \bar{\mathcal H}.
$
{
In this notation, left- or right-multiplication become ordinary operators acting on the two factors of
$\mathcal H\otimes \bar{\mathcal H}$, so $
|AXB\rangle\!\rangle = (A\otimes B^T)\,|X\rangle\!\rangle$.} 
Hence the superoperator $\ad \, A$ becomes an ordinary operator on the doubled Hilbert space:
\begin{equation}
\ad \, A = A\otimes \mathbf 1 - \mathbf 1\otimes A^T.
\end{equation}
Therefore,
\begin{equation}
\label{eq:ad_to_fund}
\Tr_{\ad}\prod_{m=1}^{10}(\ad \, Y_m) 
=
\Tr 
\prod_{m=1}^{10}\bigl(Y_m\otimes \mathbf 1-\mathbf 1\otimes Y_m^T\bigr)
= \sum_S (-1)^{|S|}\,
\Tr \Big[\prod_{m\notin S}^{\rightarrow} Y_m\Big]\,
\Tr\Big[\prod_{m\in S}^{\rightarrow} Y_m^T\Big].
\end{equation}
Here the sum runs over subsets $S\subset\{1,\ldots,10\}$.
Using the fact that the transpose reverses the order of matrix multiplication,
one obtains the desired factorization:
\begin{equation}
\Tr_{\ad}\prod_{m=1}^{10}(\ad \, Y_m)
= \sum_S (-1)^{|S|}\,
\tr\Big[\prod_{m\notin S}^{\rightarrow}Y_m\Big]\,
\tr\Big[\prod_{m\in S}^{\leftarrow}Y_m\Big].
\end{equation}
So the adjoint trace is automatically a sum of single-trace and double-trace
terms. There are no additional hidden powers of $N$ in the conversion from
adjoint to fundamental traces. 

Now, for the special case in the sum where the set $S$ is the empty set or the full set, we get a factor of  $\tr(\mathbf 1)=N$ multiplying a single trace; in all other cases we get a double trace.  
Therefore $\mathcal{O}_{10}$ decomposes as
\begin{equation}
\label{eq:O10_largeN_decomp}
\mathcal{O}_{10}
= 2N\,\mathcal{T}_{10}
+ N^2\sum_{p=2}^{8}\mathcal{D}_p,
\end{equation}
where we defined the single-trace pseudoscalar
\begin{equation}
\label{eq:T10def}
\mathcal{T}_{10}
= \epsilon_{i_1\cdots i_9}\,
\tr(Y_0 Y_{i_1}\cdots Y_{i_9}),
\end{equation}
with an unnormalized fundamental trace. With this convention
$\ev{\mathcal T_{10}^2}$ is $O(N^0)$ in the planar limit.
The operators $\mathcal{D}_p$ ($p=2,\ldots,8$) denote the normalized
double-trace contributions from subsets with $|S|=p$:
\[
\mathcal{D}_p
= \sum_{\substack{S\subset\{0,\ldots,9\}\\|S|=p}}
(-1)^{p}\,
\epsilon_{i_1\cdots i_9}\,
\left(\frac{1}{N}\tr\prod_{k\notin S}^{\to}\! Y_k\right)
\left(\frac{1}{N}\tr\prod_{k\in S}^{\leftarrow}\! Y_k\right).
\]
Note that the $p=1,9$ terms vanish because the matrices are traceless.

\subsection{Large \texorpdfstring{$N$}{N} factorization}
\label{sec:factorization} 
We now show that the single-trace piece controls the leading
$N^2$ scaling of $\ev{\mathcal{O}_{10}^2}$, {which dictates the leading contribution to $\ev{\theta^2}$}.
First note that $\mathcal T_{10}$ and each $\mathcal D_p$ are linear in each of the
ten matrices $Y_0,\ldots,Y_9$. The bosonic zero-mode measure is invariant
under flipping the sign of any one component, while these operators change
sign under such a reflection. Therefore
\begin{equation}
\label{eq:onept_vanish}
\ev{\mathcal{T}_{10}} = 0,\qquad
\ev{\mathcal{D}_p} = 0.
\end{equation}
Now in the 't Hooft limit, fully connected correlators of normalized traces %
$\tau_i=\frac{1}{N}\tr W_i(Y)$ %
scale as
\begin{equation}
\label{eq:factorization_rule}
\ev{\tau_1\cdots\tau_m}_c \propto N^{2-2m}.
\end{equation}
This implies that the three kinds of terms in
\begin{align}
\ev{\mathcal{O}_{10}^2}
= 4N^2\ev{\mathcal{T}_{10}^2}
+ 4N^3\sum_{p=2}^{8}\ev{\mathcal{T}_{10}\mathcal{D}_p}
+ N^4\sum_{p,q=2}^{8}\ev{\mathcal{D}_p\mathcal{D}_q}
\end{align}
scale as follows.
$\ev{\mathcal{T}_{10}^2} \propto  N^{0}$, since $\mathcal T_{10}$ is $N$ times a normalized single trace.
The cross term $\ev{\mathcal{T}_{10}\mathcal{D}_p} \propto N^{-3}$. This is because the disconnected piece vanishes $\ev{\mathcal{T}_{10}}\ev{\mathcal{D}_p} = 0 $ by~\eqref{eq:onept_vanish}, leaving $N$ times the connected three-point of normalized single traces, which is obtained by setting $m=3$ in \eqref{eq:factorization_rule}.
The double-trace term  $\ev{\mathcal{D}_p\mathcal{D}_q} \propto  N^{-4}$; again $\ev{\mathcal{D}_p}=0$ kills all disconnected pieces, and the leading contribution comes from the two pair-connected two-point functions of the four underlying normalized single traces, $N^{-2} \cdot N^{-2} \propto N^{-4}$. The connected four-point piece is $\propto N^{-6}$, and is subleading.
Therefore
\begin{equation}
\label{eq:O10sq_largeN}
\ev{\mathcal{O}_{10}^2}
= 4N^2\,\ev{\mathcal{T}_{10}^2}+O(N^0).
\end{equation}
The counting is
$\mathcal O_{10}\sim 2N\mathcal T_{10}$ and
$\ev{\mathcal T_{10}^2}_c\sim N^{0}$, so
$\ev{\mathcal O_{10}^2}\sim (2N)^2=4N^2$.
So, the single-trace piece determines the leading $N^2$ coefficient,
while the double-trace sector first contributes at order~$N^0$. Large $N$ factorization further implies that higher moments obey Wick's theorem, implying the exponentiation claimed in \eqref{eq:final}.

Substituting~\eqref{eq:O10sq_largeN} into \eqref{eq:final} then gives
\begin{equation}
\label{eq:finalN}
\ev{\cos\theta}
= \exp\!\left(
- 2\cten^2\,N^2\,(\lambda\beta^3)^5\,
\ev{\mathcal{T}_{10}^2}
+ O\!\left(N^0(\lambda\beta^3)^5\right)
+ O\!\left((\lambda\beta^3)^{11/2}\right)
\right).
\end{equation}
Since $\ev{(\mathcal{T}_{10})^2 } \propto N^{0}$,
$\ev{\cos\theta}
\sim 
\exp \bigl(-\text{const}\cdot N^2\,(\lambda\beta^3)^5\bigr)$, as we claimed in \eqref{eqIntro1}.

\subsection{Large \texorpdfstring{$D$}{D} expansion and Monte Carlo}
\label{sec: main text large D}

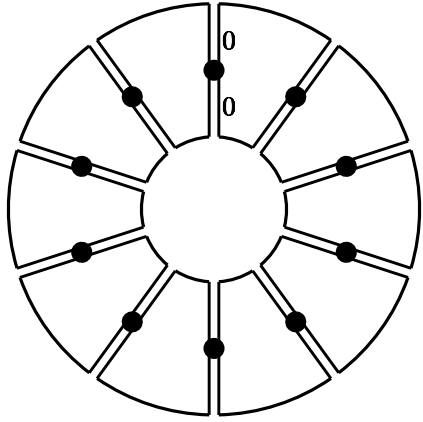
\begin{figure}[t]
\centering
\begin{tikzpicture}[scale=.8,baseline={(0,0)}]
\def\Rout{3.6}\def\Rin{3.4}\def\rout{1.2}\def\rin{1.0}
    \def\angOut{1.35}
    \def\angIn{3.85}
    \def\bone{1.7}
    \def\btwo{2.9}
    \foreach \a in {0,36,72,108,144,180,216,252,288,324} {
        \draw[black,line width=1.2pt] ({90+\a+\angOut}:\Rin) arc ({90+\a+\angOut}:{90+\a+36-\angOut}:\Rin);
        \draw[black,line width=1.2pt] ({90+\a+\angIn}:\rout) arc ({90+\a+\angIn}:{90+\a+36-\angIn}:\rout);
        \node at (0.25,1.7) {\footnotesize{0} };
        \node at (0.25,2.8) {\footnotesize{0} };
        \begin{scope}[rotate around={\a:(0,0)}]
            \draw[black,line width=1.1pt] (-0.08,\Rin) -- (-0.08,\rout);
            \draw[black,line width=1.1pt] (0.08,\Rin) -- (0.08,\rout);
            \draw[black,line width=1.1pt,fill=black] (0,{0.5*(\Rin+\rout)}) circle (0.15);
        \end{scope}
    }
\end{tikzpicture}

\caption{The planar Wick contraction of
$\ev{W_{\mathrm{id}}\,W_\rho}_0$. The rungs that connect the inner and outer ring are the dressed propagators in the large $D$ approximation for the bosonic zero modes.
}
\label{fig:planar_annulus}
\end{figure}

In this section, we will estimate the expectation value of $\mathcal{T}_{10}^2$ in the matrix model integral \eqref{eq:bIKKT}. This model can be viewed as part of a family of models with $D$ matrices. In the large $D$ limit, the model becomes solvable \cite{Hotta:1998en, Mandal:2009vz, Li:2025tub}.

As we explained, for the leading large $N$ contribution to the phase, it is enough to compute
the single-trace correlator in \eqref{eq:O10sq_largeN}.  At large $D$, one finds that the dressed propagators
\footnote{
We can sum bubble diagrams using the Schwinger-Dyson equation:
\begin{align}
\begin{tikzpicture}[>=stealth,baseline={([yshift=-.5ex] current bounding box.center)}]
 \draw[Xdouble](-2,0)--(0,0);
  \node at (-2.25,0) {$\mu$};
 \node at (0.25,0) {$\mu$};
\end{tikzpicture} \quad &= \quad 
\begin{tikzpicture}[>=stealth,baseline={([yshift=-.5ex] current bounding box.center)}]
 \draw[ydouble](-2,0)--(0,0);
\end{tikzpicture} \quad +\quad 
\begin{tikzpicture}[>=stealth,baseline={([yshift=-2.5ex] current bounding box.center)}]
 \draw[ydouble](-2,0)--(0,0);
\draw[Xdouble] (0,0.5) circle (0.5);
\draw[Xdouble](0,0)--(1.5,0);
\node at (0,.5) {$\nu$};
\node at (1.75,0) {$\mu$};
\node at (-2.25,0) {$\mu$};
\fill[blue] (0,0) circle (0.1);
\end{tikzpicture}\\
G& =  \frac{1}{m^2}-\frac{1}{m^2}\Delta^2 G, \quad \Delta^2 = 2 \frac{\lambda D}{\beta}   G  .
\end{align}
Here we have introduced a small mass term $m$ to give the bosonic fields a propagator. Then we may solve the above equations which yields $ G = \frac{1}{\Delta^2 + m^2}$. Setting $m \to 0$, 
\begin{align}
    G = \frac{1}{\Delta^2}= \frac{1}{(2 \lambda D/\beta  )^{1/2}}.
\end{align}
This is the 2-pt function of the zero mode fields $X$; after a rescaling \eqref{eq:Ydef} we arrive at \eqref{YYprop}.} acquire an effective mass \cite{Hotta:1998en}, 
\begin{equation}\label{YYprop}
\ev{Y_\mu^a Y_\nu^b}_0
=
\frac{\delta_{\mu\nu}\delta^{ab}}{N\sqrt{2D}}.
\end{equation}
Now, it is convenient to rewrite \eqref{eq:T10def} as
\begin{equation}
\mathcal{T}_{10}
=
\sum_{\sigma\in S_9}\operatorname{sgn}(\sigma)\,W_\sigma,
\quad
W_\sigma:=\tr(Y_0Y_{\sigma(1)}\cdots Y_{\sigma(9)}).
\end{equation}
Since each label $Y_0,\ldots,Y_9$ appears exactly once in each trace, every
nonzero Wick contraction pairs identical labels across the two traces.  For
fixed $\sigma$, the unique planar contraction is of 
$W_\sigma$ with $W_{\rho \cdot  \sigma}$, where $\rho$ reverses the order:
\begin{align}
\rho(1,2,\ldots,9)=(9,8,\ldots,1),
\end{align}
as shown in Fig.~\ref{fig:planar_annulus}.  
The planar annulus has ten propagators, each contributing
$1/(N\sqrt{2D})$, and ten index loops, contributing $N^{10}$.  Therefore
\begin{equation}
\ev{W_\sigma W_{\rho\cdot\sigma}}_0
=
\frac{N^{10}}{(N\sqrt{2D})^{10}}
=
\frac{1}{(2D)^5}
\end{equation}
There are exactly $9!$ ordered planar pairs $(\sigma,\rho\circ\sigma)$ in
$\ev{\mathcal T_{10}^2}_0$. Using $\operatorname{sgn}(\rho)=1$, this gives
\begin{equation}
\ev{\mathcal{T}_{10}^2}_0
=
\frac{9!}{(2D)^5}
+ O(N^{-2}).
\end{equation}
Substituting the leading large-$D$ result into~\eqref{eq:finalN} then gives the
explicit large-$N$ exponent
\begin{equation}
\label{eq:cos_largeD}
\ev{\cos\theta}_{0}
=
\exp\!\left[
-c \,
N^2(\lambda\beta^3)^5
\right], \\
\quad c = \frac{2\cten^2\,9!}{(2D)^5}\left(1 + \frac{7}{3D} + \cdots\right).
\end{equation}
Here we have also included the first subleading $1/D$ correction, which we explain in Appendix \ref{largeDexpansion}. 
For the BFSS value $D=10$, this yields $c \simeq 8.2\times 10^{-9}$. %
We see that the small numerical value of $c$ is related to the fact that $D=10$ is a relatively large number. It is amusing to note that this numerology traces back to the fact that string theory is a theory in 10 dimensions.%

As an alternative to the large $D$ expansion, we also performed a Monte Carlo simulation of the bosonic zero-mode theory and used it to compute $c$, with approximate agreement with the large $D$ expansion, see Appendix \ref{MC}. This involved performing a large $N$ extrapolation; for numerical stability we also found it convenient to add a small mass term and performed a small-mass extrapolation as well. This gave an estimate
\begin{align}
    \ev{(\mathcal T_{10})^2} \simeq 0.156, \quad c \simeq 9.1 \times 10^{-9}.
\end{align}
This number comes with some systematic and statistical uncertainties which we discuss in Appendix \ref{MC}. From the large $D$ expansion, we expect that subleading terms in $1/D$ could correct the large $D$ estimate by $\sim 20\%$ so the two methods are in reasonable agreement.

\section{The replica method}
\label{sec:replica}

We can also obtain the leading operator from a diagrammatic expansion of
the fermion determinant in the high-temperature limit. This provides a
cross-check on the operator content of section~\ref{sec:hightemp} and
makes the exponentiation of $\log\ev{\cos\theta}$ at large~$N$ manifest. 
This formalism enables us to study the sign problem using conventional planar Feynman diagrams, albeit with a replica trick.

\subsection{Replica formulation of the phase-quenched theory}
\label{sec:replica_formulation}

Let us define
the following path integrals  %
\begin{align}
Z_\bfss^{(2n)}
&=\int\D X\,\Pf(\mathcal M)^{2n}\,e^{-S_b},\quad
Z_\qu^{(2n)}
=\int\D X\,\bigl(\Pf(\mathcal M)\Pf(\mathcal M)^*\bigr)^n\,e^{-S_b}.
\end{align}
For integer $n$, the second quantity 
admits a realization with $n$ replicas of
the original fermion $\psi^A$ ($A=1,\dots,n$) and $n$ replicas of a conjugate
fermion $\chi^A$,
\begin{align}
\label{eq:Sabs_replica}
    Z_\qu^{(2n)}
&=\int\D X\,\D \psi^A\, \D \chi^A\,e^{-S^{(2n)}_\qu},\\
\nonumber
S^{(2n)}_\qu
&= \frac{N}{\lambda} \!\int_0^\beta\!\dd t\,\tr\Bigg[
\frac12 (D_t X_i)^2 - \frac14 [X_i,X_j]^2
\\
&+\sum_{A=1}^n\Big(\tfrac12 \psi^A_\alpha D_t \psi^A_\alpha
- \tfrac12\psi^A_\alpha (\gamma_i)_{\alpha\beta}[X_i,\psi^A_\beta]\Big)
+\sum_{A=1}^n\Big(\tfrac12 \chi^A_\alpha D_t \chi^A_\alpha
+ \tfrac12\chi^A_\alpha (\gamma_i)_{\alpha\beta}[X_i,\chi^A_\beta]\Big)
\Bigg].
\end{align}
We will refer to $\chi$ as ``conjugate fermions.'' Meanwhile, $Z_\bfss^{(2n)}$ admits a realization in terms of $2n$ replicas of the original fermion $\psi^A$. The idea is that integrating out $\psi^A$ gives $\Pf(\mathcal M)^n$ and integrating
out $\chi^A$ gives $\Pf(\mathcal M^*)^n$, so that
\begin{equation}
\label{eq:replica_id}
\frac{Z_\bfss^{(2n)}}{Z^{(2n)}_\qu}
=\frac{\int\!\D X\,\Pf(\mathcal M)^{2n}\,e^{-S_b}}
       {\int\!\D X\,|\Pf(\mathcal M)|^{2n}\,e^{-S_b}},
\qquad
\ev{\cos\theta} =
\lim_{n\to 1/2}\frac{Z_\bfss^{(2n)}}{Z^{(2n)}_\qu}.
\end{equation}
The last equality uses the reflection symmetry of the zero-mode measure,
which sets $\ev{\sin\theta}=0$ as in section~\ref{sec:hightemp}.

Note that the sign in the
$\chi$ Yukawa coupling is fixed by complex conjugating the adjoint-basis
kernel.  Since $(\ad \, Y)^*=-\ad \, Y$, the term $-\ii\ad \, A$ is unchanged under
complex conjugation, while the spatial Yukawa piece $-\ad \, X\gamma_i$ changes
sign.  Equivalently, after the harmless Matsubara relabeling
$r\mapsto-r$, the local kernel for $\mathcal M^*$ is obtained from the
original one by keeping $D_t$ and sending $X_i\to-X_i$ in the fermion
coupling. Then, with both $\psi$ and $\chi$ transforming in the adjoint, the
replica action above is gauge invariant and integrates to
$\Pf(\mathcal M)^n\Pf(\mathcal M^*)^n$.

 At integer
$n$ both partition functions are ordinary local fermionic replica theories; the logarithm
of each replicated partition function is computed by connected vacuum
diagrams in the corresponding local theory. We consider these next, to compute $\ev{\cos\theta}$.

\subsection{Cancellation between \texorpdfstring{$\psi^A$}{psi\textasciicircum A} and \texorpdfstring{$\chi^A$}{chi\textasciicircum A} loops and the leading diagram}
\label{sec:replica_cancellation}
We now work out the Feynman rules for the replica theories in Fourier space. We denote the free fermion propagators 
\begin{equation}
\begin{tikzpicture}[scale=0.5,line cap=round,line width=1.5pt,baseline=0.20cm]
\draw[blue] (0,1) -- (7,1);
\draw[blue] (0,0.25) -- (7,0.25);
\end{tikzpicture}
\quad = \big\langle\psi^A_{r,\alpha}\,\psi^B_{-r',\beta}\big\rangle
=
\frac{\delta^{AB}\,\delta_{rr'}\,\delta_{\alpha\beta}}{2\pi\ii r},
\end{equation}
\begin{equation}
\begin{tikzpicture}[scale=0.5,line cap=round,line width=1.5pt,baseline=0.20cm]
\draw[red] (0,1) -- (7,1);
\draw[red] (0,0.25) -- (7,0.25);
\end{tikzpicture}
\quad = \big\langle\chi^A_{r,\alpha}\,\chi^B_{-r',\beta}\big\rangle
=
\frac{\delta^{AB}\,\delta_{rr'}\,\delta_{\alpha\beta}}{2\pi\ii r}.
\end{equation}
On the static background of section~\ref{sec:hightemp}, the fermionic action consists of the two interaction terms
\begin{align}
\varepsilon &= (\lambda \beta^3)^{1/4} \\
   V_{i,\psi} &= \frac12 \varepsilon  N \sum_r  \Tr \,  \psi^A_{-r,\alpha} (\gamma_i)_{\alpha\beta}[Y_i,\phi^A_{r,\beta}] =   \varepsilon  N  \sum_r \Tr\,   \psi^A_{r,\alpha}(\gamma_i)_{\alpha\beta}Y_i\psi^A_{r,\beta} \\
   V_{0,\psi} &= \frac{\ii}{2} \varepsilon  N \sum_r  \Tr \,  \psi^A_{-r,\alpha} [Y_0,\phi^A_{r,\beta}] =  \ii  \varepsilon  N  \sum_r \Tr\,   \psi^A_{r,\alpha}Y_0\psi^A_{r,\beta} \\
   V_{i,\chi} &= -   \varepsilon  N  \sum_r \Tr\,   \chi^A_{r,\alpha}(\gamma_i)_{\alpha\beta}Y_i\chi^A_{r,\beta} \\
   V_{0,\chi} &= \ii  \varepsilon  N  \sum_r \Tr\,   \chi^A_{r,\alpha}Y_0 \chi^A_{r,\beta}
\end{align}
corresponding to two kinds of interaction insertions on a fermion line. Vertices of the first kind are represented as
\begin{align}
    \begin{tikzpicture}[scale=0.5,line cap=round,line width=1.5pt,baseline=(current bounding box.center)]
\draw[blue] (0,3) -- (7,3);
\draw[blue] (0,2.25) -- (3.125,2.25);
\draw[blue] (3.875,2.25) -- (7,2.25);
\draw[black] (3.125,2.25) -- (3.125,0.5);
\draw[black] (3.875,2.25) -- (3.875,0.5);
\end{tikzpicture} \quad &= \quad  +\varepsilon \gamma_i \otimes  Y^i\\
 \begin{tikzpicture}[scale=0.5,line cap=round,line width=1.5pt,baseline=(current bounding box.center)]
\draw[red] (0,3) -- (7,3);
\draw[red] (0,2.25) -- (3.125,2.25);
\draw[red] (3.875,2.25) -- (7,2.25);
\draw[black] (3.125,2.25) -- (3.125,0.5);
\draw[black] (3.875,2.25) -- (3.875,0.5);
\end{tikzpicture} \quad &= \quad  -\varepsilon \gamma_i \otimes Y^i.
\end{align}
For the vertex with $Y_0$ we instead replace $\gamma_i \to \ii \cdot \mathbf{1}$. We will distinguish these vertices involving the gauge field $Y_0$ by writing $0$ next to the vertex:
\begin{align}
    \begin{tikzpicture}[scale=0.5,line cap=round,line width=1.5pt,baseline=(current bounding box.center)]
\draw[blue] (0,3) -- (7,3);
\draw[blue] (0,2.25) -- (3.125,2.25);
\draw[blue] (3.875,2.25) -- (7,2.25);
\draw[black] (3.125,2.25) -- (3.125,0.5);
\draw[black] (3.875,2.25) -- (3.875,0.5);
\node at (4.4,1) {0};
\end{tikzpicture} \quad &= \quad      \begin{tikzpicture}[scale=0.5,line cap=round,line width=1.5pt,baseline=(current bounding box.center)]
\draw[red] (0,3) -- (7,3);
\draw[red] (0,2.25) -- (3.125,2.25);
\draw[red] (3.875,2.25) -- (7,2.25);
\draw[black] (3.125,2.25) -- (3.125,0.5);
\node at (4.4,1) {0};
\draw[black] (3.875,2.25) -- (3.875,0.5);
\end{tikzpicture} \hspace{-1.5cm} &= \quad \varepsilon  \ii \mathbf{1} \otimes  Y^0
\end{align}

{We first consider diagrams for the partition function that involve a single fermion ring.} A single connected fermion ring with $p$ insertions of $V_0$ and $q$
insertions of the spatial vertex $V_i$ contributes
\begin{equation}
\label{eq:replica_loop_count}
\log Z^{(2n)}_\qu\supset n\bigl(1+(-1)^q\bigr)\,\Lambda_{p,q},
\qquad
\log Z_\bfss^{(2n)}\supset 2n\,\Lambda_{p,q},
\end{equation}
where $\Lambda_{p,q}$ is the diagram value with a single fermion species.
The single ring contribution to the difference
$\log Z_\bfss^{(2n)}-\log Z^{(2n)}_\qu$ is therefore
$n(1-(-1)^q)\Lambda_{p,q}$, which vanishes for even~$q$. When $q$ is odd, a
non-zero Matsubara sum over $r$ requires $p+q$ to be even, and therefore $p$ must also
be odd. The same gamma-trace selection rule of section~\ref{sec:hightemp}
then applies: the trace of fewer than nine $\gamma$ matrices vanishes in the
16-dimensional Clifford representation. Therefore, the leading fermion ring that contributes has $(p,q)=(1,9)$ and is of order $(\lambda\beta^3)^{5/2}$.

\begin{figure}[t]
\centering
\begin{tikzpicture}[scale=.8,baseline={(0,0)}]
    \def\Rout{3.6}\def\Rin{3.4}\def\rout{1.2}\def\rin{1.0}
    \def\angOut{1.35}\def\angIn{3.85}
    \def\bone{1.7}\def\btwo{2.9}
    \def\angBtwo{1.6}\def\angBone{2.7}

    \draw[blue,line width=1.2pt] (0,0) circle (\Rout);
    \node at (0.25,1.45) {\footnotesize{0} };
    \node at (0.25,3.15) {\footnotesize{0} };
    \foreach \a in {0,36,72,108,144,180,216,252,288,324} {
        \draw[blue,line width=1.2pt] ({90+\a+\angOut}:\Rin) arc ({90+\a+\angOut}:{90+\a+36-\angOut}:\Rin);
        \draw[red,line width=1.2pt] ({90+\a+\angIn}:\rout) arc ({90+\a+\angIn}:{90+\a+36-\angIn}:\rout);
        \begin{scope}[rotate around={\a:(0,0)}]
            \draw[black,line width=1.1pt] (-0.08,\Rin) -- (-0.08,\btwo);
            \draw[black,line width=1.1pt] (0.08,\Rin) -- (0.08,\btwo);
            \draw[black,line width=1.1pt] (-0.08,\rout) -- (-0.08,\bone);
            \draw[black,line width=1.1pt] (0.08,\rout) -- (0.08,\bone);
        \end{scope}
    }
    \draw[red,line width=1.2pt] (0,0) circle (\rin);
    \node[above=2pt,color=blue] at (0,\Rout) {$\psi_A$};
    \node[above=2pt,color=red] at (0,-\rin) {$\chi_B$};
            \fill[even odd rule,pattern={Lines[angle=45,distance=5pt,line width=1pt]},pattern color=gray] (0,0) circle (\btwo) (0,0) circle (\bone);
    \foreach \a in {0,36,72,108,144,180,216,252,288,324} {
        \draw[black,line width=1.1pt]
            ({90+\a+\angBtwo}:\btwo)
            arc ({90+\a+\angBtwo}:{90+\a+36-\angBtwo}:\btwo);
        \draw[black,line width=1.1pt]
            ({90+\a+\angBone}:\bone)
            arc ({90+\a+\angBone}:{90+\a+36-\angBone}:\bone);
    }
\end{tikzpicture}

\caption{%
The leading planar diagrams that contribute to $\ev{\cos \theta}$ in the replica expansion. The bosonic rungs enter the annulus
from the outer fermion ring and are absorbed into the non-perturbative
zero-mode correlator. This presentation makes manifest that only the external
bosonic legs that contract the two single-trace operators
$\mathcal T_{10}$ are kept explicit; the resummation of all bosonic
interactions inside the connected two-point function lives entirely
inside the hashed annulus. We may think of Figure \ref{fig:planar_annulus} as a possible contribution to the annulus. }
\label{fig:replica_nonperturbative_internal}
\end{figure}
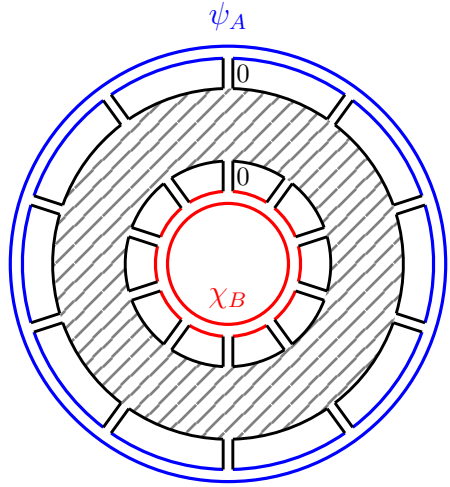

This leading one-ring structure with $(p,q)=(1,9)$ is precisely the pseudoscalar operator
$\mathcal T_{10}$ found in the direct high temperature expansion. Its
bosonic one-point function vanishes by the reflection symmetry of the
zero-mode measure. More generally, in a connected graph with several closed
fermion rings, the replica species factor in $ \log Z_\bfss^{(2n)}$ is
$(2n)^L$, while the corresponding factor in $\log Z^{(2n)}_\qu$ is $n^L\prod_{\ell=1}^L\bigl(1+(-1)^{q_\ell}\bigr)$, where the first term comes from a $\psi$ ring and the second factor comes from a $\chi$ ring with $q_\ell$ insertions of $V_i$ on ring $\ell$. Hence
\begin{equation} \label{replica}
\log Z_\bfss^{(2n)} -  \log Z^{(2n)}_\qu \propto (2n)^L - n^L\prod_{\ell=1}^L\bigl(1+(-1)^{q_\ell}\bigr).
\end{equation}
If all $q_\ell$ are even, we see that the two factors cancel. If an odd number of loops have
odd $q_\ell$, the graph contains an odd total number of spatial $Y_i$
insertions and vanishes in the bosonic zero-mode average. The first non-vanishing
contribution to the difference above therefore comes from two
fermion rings, with each ring carrying $(p,q)=(1,9)$, and
appears at order~$(\lambda\beta^3)^5$. This corresponds to the contribution $\ev{\mathcal O}_{10}^2$ derived in the high temperature expansion before. %

Let us now evaluate the leading planar diagram, see Figure \ref{fig:replica_nonperturbative_internal}.
 Each diagram contains 20 vertices, with 10 propagators on the inner circle and 10 on the outer circle. The two $V_0$ vertices also come with a factor of $\ii^2$. Additionally, we also get a factor of $N^2$ from the outermost and innermost fermionic loop in the 't Hooft double notation. Combining everything, this gives
\begin{align}
    &\ev{\ii^2  N^2 \varepsilon^{20}  \left[S_{10}\Tr(\gamma^{i_1} \cdots \gamma^{i_9})  \Tr(Y^0 Y^{i_1} \cdots Y^{i_9} ) \right]^2} = -\varepsilon^{20} (16 S_{10})^2  N^2 \ev{(\mathcal{T}_{10})^2} \\
     S_{10} &=   \sum_{r\in\mathbb Z+\frac12}
    \frac{1}{(2\pi \ii r)^{10}}
    =
    -\frac{31}{1451520}.
\end{align}
Thus the phase coefficient defined in \eqref{eq:final} obeys
\(\cten=-8S_{10}\).
Here we have included the factor
$\left[\Tr(\gamma^{i_1} \cdots \gamma^{i_9})  \Tr(Y^0 Y^{i_1} \cdots Y^{i_9} ) \right]$ for each ring.
Now the full contribution should include the replica factor \eqref{replica} as well as a symmetry factor of 1/2. This symmetry factor is that of the uncolored diagram\footnote{Imagine contributions to either the BFSS$^{(2n)}$ theory or the sign-quenched replica theory. There will be connected Feynman diagrams where the inner and outer ring have the same species of fermion. The factor of 1/2 is the correct symmetry factor for such diagrams. When they are different, there is an overcounting by a factor of 2 in \eqref{replica} which is remedied by the factor of 1/2.}, which has a $Z_2$ symmetry of swapping the inner and outer ring. This leads to
\begin{align}
\log Z_\bfss^{(2n)} - \log Z^{(2n)}_\qu  &= -\frac12   \times (2n)^2 \times   \varepsilon^{20} (16 S_{10})^2  N^2 \ev{(\mathcal{T}_{10})^2} %
\end{align}
Now we continue to $n \to 1/2$, which gives
\begin{align}
\log Z_\bfss - \log Z_\qu  &= -2 \cten^2    \varepsilon^{20}  N^2 \ev{(\mathcal{T}_{10})^2}, %
\end{align}
in agreement with \eqref{eq:finalN}.

\section{The sign problem in higher dimensional SYM} \label{sign_higherD}
A natural generalization of the BFSS problem is to consider maximally supersymmetric Yang Mills theories in $d$ spacetime dimensions on $\mathcal{M}_{d-1} \times S_1$ with thermal boundary conditions on the $S_1$.\footnote{One can also consider supersymmetric boundary conditions on the $S_1$, in which case one reduces to a lower dimensional SYM theory. For example, in $d=2$, one can put the theory on a torus and consider a regime where one cycle is small. If we impose supersymmetric boundary conditions on this small cycle, we reduce to BFSS. If the long cycle has thermal boundary conditions, our BFSS analysis applies. The new case that we analyze in this section is when the short cycle has thermal boundary conditions. The long cycle may have either periodic or anti-periodic boundary conditions for the fermions.} 
Monte Carlo simulations for higher-dimensional maximally supersymmetric SYM as well as their less supersymmetric cousins have been studied on the lattice in a range of works, see \cite{Catterall:2010fx,Hanada:2010qg,Mehta:2011ud,Galvez:2012sv,Catterall:2012yq,Kadoh:2017mcj,Catterall:2017lub,Catterall:2017xox,Sherletov:2022rnl,Sherletov:2023udh,Catterall:2023tmr,Joseph:2026fdm}. Several authors (see in particular \cite{Galvez:2012sv, Catterall:2010fx} for the 1+1D $\mathcal{N} =$(8,8) SYM case)  comment on the mildness of the sign problem for the maximally supersymmetric case as the continuum limit is approached. %

In principle, one could follow the steps that we performed above in the BFSS case for the $d$-dimensional SYM theory on $\mathbb{R}^{d-1} \times S_1$. This is done for $d=2$ in Appendix \ref{app:higherD}.
Upon dimensional reduction with anti-periodic boundary conditions for the fermions, we obtain a $d-1$ dimensional bosonic Yang-Mills theory with adjoint matter which has an enhanced global symmetry group. One then analyzes the transformation of the phase under the relevant discrete symmetries and identifies the leading local operator in the reduced theory consistent with these symmetries. On general EFT grounds, we expect that the phase should be an integral over the $d-1$ dimensions of a {\it local operator} in the high temperature limit, since upon Kaluza-Klein compactification the fermions are very massive and integrating them out should lead to a local operator.  The perturbative high-temperature expansion should then give an expansion in terms of derivatives of local operators. Of course computing the coefficients of these operators will be more laborious in higher dimensions.

In this section, we will instead treat this problem somewhat heuristically by taking a shortcut. The higher dimensional SYM theories can be obtained from BFSS by considering special large $N$ matrix configurations and expanding the action around these configurations \cite{Taylor:1996ik}. We can then estimate what operators could appear in higher $d$ by plugging in these special large $N$ matrix configurations into our formula for the BFSS Pfaffian. Let us illustrate this for $d=2$, which is the $\mathcal{N}=(8,8)$ SYM theory. This consists of two steps:
 \begin{align}
    \text{BFSS} \to \text{1+1D SYM} \to \text{bosonic BFSS}.
\end{align}
In the first step, we uplift to 1+1D SYM by 
 considering two infinite dimensional matrices $x,p$ which satisfy $[x,p] = \ii $. %
 We set
\begin{align}
    X_9 = p_x-A_x(x,t), \quad X_i = X_i(x,t), \quad A_0 = A_0(\tau,x).
\end{align}
In the second step, we dimensionally reduce 1+1D SYM on a thermal circle with anti-periodic boundary conditions for the fermions. This can be done easily using our existing formulas for $\theta$ as an operator, which hold for arbitrary matrix configurations. This gives us an expression $\theta = \int \d x \, O(x)  $ where $O(x)$ is a local operator. This should be interpreted as an expression in the bosonic BFSS theory, e.g., a quantum mechanics with the same bosonic adjoint fields as BFSS but with no fermions. It is a Euclidean theory, since it represents the spatial circle of the 1+1D SYM theory. In this dimensionally reduced theory, the original BFSS gauge field $A_0$  gets reinterpreted in the bosonic BFSS model as the ninth matrix $A_0\to X_9$, and $A_x$ (the gauge field in the 1+1d SYM theory) becomes the gauge field $A_0$ in the bosonic BFSS theory.

More precisely, using the coefficient $\cten$ defined in \eqref{eq:final}, we have
\begin{align} \label{thetaD}
    \theta &\simeq  2 \cten \, N\,  (\lambda \beta^3)^{5/2} \,\mathcal{T}_{10} \\
    &=  2 \cten  \, N\, \beta^{10}\epsilon_{i_1\cdots i_8 }  \sum_{m=0}^8 (-1)^m   \,
\tr \left({A_0} {X_{i_1}}\cdots X_{i_m} X_{9} X_{i_{m+1}}\cdots X_{i_8} \right)\\
    &=  2\cten N \beta^{10}  \epsilon_{i_1\cdots i_8 } \bigg\{\sum_{\text{odd } m =1 }^7  \big[ 
\tr \left({A_0} {X_{i_1}}\cdots [X_9, X_{i_m} ] \cdots  X_{i_8} \right) \big] + \Tr \left({A_0} {X_{i_1}}\cdots  X_{i_8} X_9 \right) \bigg\}%
\end{align}
The trace over the SU(N) indices becomes an integral over the worldvolume $x$ coordinate, giving rise to an integral over a local operator in the lower dimensional ``bosonic BFSS'' theory:
\begin{align} \label{theta2D}
    \theta_{d=2} = 2\cten N \beta^{10}\int \d x \, \epsilon_{i_1 \cdots i_8} \sum_{\ell=0}^3 \tr \left({X_9} {X_{i_1}}\cdots (D_x X_{i_{2\ell+1}})  \cdots  X_{i_8} \right) + \cdots,
\end{align}
where we have dropped a singular term and relabeled $A_0 \to X_9$. 
This lower dimensional theory has $\lambda' = \lambda_{d=2} / \beta$ and the $X$ or $A_x$ fields have mass dimension 1, so $\theta^2 \sim N^2  \beta^{20} \ev{X^{20}} \sim N^2\beta^{20} (\lambda')^{20/3} \sim N^2  \beta^{40/3} \lambda_{d=2}^{20/3}$.%

The above analysis \eqref{thetaD} is schematic for the following reason. It treats the spatial momentum $p$ in $D_x$ as an $O(1)$ variable, and then expands $\Tr \log \mathcal{M}$ in $\beta$, holding $p$ fixed. However, $p$ is integrated over in the fermion propagators. We should instead use the propagators for a tower of Kaluza-Klein fermions in $d-1$ dimensions $1/(\slashed{p} - m_r)$ where $m_r=2\pi r/\beta$. When we do so, the typical momentum in the integral is $p\sim 1/\beta$, technically invalidating the expansion \eqref{thetaD}. Nevertheless, the more careful treatment in Appendix \ref{app:higherD} reproduces the same operator without the singular term and including the correct coefficients.

More generally if we start with SYM in $d$ dimensions, repeating these arguments gives a scaling
\begin{align}
    \ev{\theta^2} \propto N^2 \beta^{20-20/(5-d)} \lambda_d^{20/(5-d)}.
\end{align}
Note that for $d=4$, the famous case of $\mathcal{N}=4$ SYM, we find a sign problem that is independent of $\beta$ but suppressed at weak-coupling by $\lambda^{20}$. Note that this is for the theory on $\mathbb{R}^{3} \times S_1$; it would be interesting to study this for other cases like $S_3 \times S_1$.

\section{Discussion}
\label{sec:discussion}
\subsection{Implications for Monte Carlo}
Let us discuss briefly the implications of our findings for Monte Carlo simulations. The small coefficient in \eqref{eqIntro1} explains why previous studies found a value of $\ev{\cos \theta} \approx 1$ for modest $N$. Even for $N=16, \beta\lambda^{1/3}=1$, we expect that $1-\ev{\cos \theta} \approx 2.3 \times 10^{-6}$. Even at $\beta \lambda^{1/3} = 2$, the exponent is still $\approx -0.076$. Although we did not explicitly compute it, we expect from the finite $N$ computation \eqref{eq:finalN} that the first non-planar correction to the exponent, of order $O(N^0)$, is suppressed by $(\lambda \beta^3)^5$.

As we mentioned in the introduction, the simplest observable of interest in Monte Carlo is the energy $E(T)$ for which there is a gravity prediction from the Bekenstein-Hawking formula for the entropy of a black hole:
\begin{align}
{\tilde{E}} = N^2 {\left(a_0 \tilde{T}^{\frac{14}{5}}+a_1 \tilde{T}^{\frac{23}{5}}+a_2 \tilde{T}^{\frac{29}{5}}+\cdots\right)} +O(N^0).
\end{align}
Here the Type IIA black hole predicts $a_0   = \frac{9}{14} 4^{13 / 5} 15^{2 / 5}(\pi / 7)^{14 / 5} \approx 7.41$  and the Monte Carlo values are $a_1 = - 9.90 \pm 0.31 $ and $a_2 = 5.78\pm 0.38$ \cite{Pateloudis:2022ijr}.

We may use our perturbative estimate to quantify the expected correction to the free energy $F(T)$ of the quenched theory relative to the BFSS model as predicted by pure IIA supergravity. To do this, we simply take the log of \eqref{eqCosTheta}:
\begin{align} 
    \log Z_\bfss - \log Z_\qu = - N^2 f(\lambda^{1/3} \beta).
\end{align}
Then we can estimate $\log Z_\bfss$ using the gravity solution \cite{Klebanov:1996un, Itzhaki:1998dd}:
\begin{align} \label{sugra_predict}
    \log Z_\text{sugra} &\simeq \frac{5}{9} a_0 N^2 \, T^{9/5} , \quad  \frac{5}{9} a_0 \approx 4.115, \\ %
    \delta F /F &\simeq {0.10}, \qquad T=0.35.
\end{align} 
Here we have used a representative value of temperature in 't Hooft units $T=0.35$ from the simulations in \cite{Pateloudis:2022ijr}. Such temperatures are not quite low enough for stringy corrections from the pure supergravity calculation to be negligible; using $N^{-2} \log Z_\qu \approx \tfrac{5}{9} a_0 T^{9/5} + \frac{5}{18} a_1 T^{18/5} +  \frac{5}{24} a_2 T^{24/5} $  we get a slightly larger estimate $\delta F/ F \sim 0.11$. 
By differentiating, we also obtain estimates  $\delta E/E \sim 1$ for $T=0.35$ (and hence the estimate may not be reliable), but $|\delta E /E| \sim 0.11$ for $T=0.4$ and $|\delta E/E| \sim 0.0030$ for $T=0.5$.
Thus we see that at temperatures $\gtrsim 0.35$ the Monte Carlo predictions for the thermodynamics should be reliable within about $\sim 10\%$. For higher temperatures, corrections due to the sign problem will be rapidly suppressed. %

It seems likely that accounting for the sign problem could lead to some shifts in the estimates of $a_1$ and $a_2$.
In the future, we imagine accounting for the difference between $\log Z_\bfss - \log Z_\qu$ by computing $\ev{\cos \theta}$ perturbatively and then using this to refine the Monte Carlo predictions for $a_1$ and $a_2$ (and perhaps this would lead to better agreement with $a_0$).

\subsection{Future directions}
An interesting question is whether the sign problem becomes less severe at intermediate or strong coupling, e.g., whether $f(\lambda^{1/3} \beta)$ is a non-monotonic function. In principle, one could at least perturbatively address this by computing higher order contributions to $f(\lambda^{1/3}\beta).$ 
At strong 't Hooft coupling, could the holographic dictionary \cite{Itzhaki:1998dd} say anything about this? If for some unknown reason $f \to 0$ at strong coupling, then one could make the interesting prediction that the bootstrap results \cite{Lin:2023owt, Lin:2024vvg} at $T=0$ should agree with the low-temperature extrapolation of Monte Carlo results. 
We have focused on the 't Hooft regime in this paper where the gravity dual is a Type IIa black hole. It would be interesting to explore the sign problem in the M-theory regime.

There are several natural directions for building on this analysis. First, it would be interesting to repeat this analysis in the BMN matrix model \cite{Berenstein:2002jq}, or other supersymmetric matrix models with a sign problem. In practice, due to challenges with the flat directions, recent Monte Carlo simulations \cite{Pateloudis:2022ijr} simulate the BMN model with finite mass deformation $\mu$ and then extrapolate to small $\mu$. 

Second, one could use these methods to estimate corrections not just to the thermodynamics but also to observables such as $\ev{\tr X^I X^I}$ in the thermal ensemble. This could be useful when one compares such observables between phase-quenched Monte Carlo and the BFSS matrix bootstrap \cite{Lin:2023owt, Lin:2024vvg}, as the bootstrap does not have a sign problem. One could wonder whether certain expectation values are more or less contaminated by the sign problem at strong coupling. The replica formulation seems particularly efficient for answering some of these questions. %

\section*{Acknowledgments}
We thank Simon Catterall, Masanori Hanada, Igor Klebanov, Juan Maldacena, Jun Nishimura, and Steve Shenker for useful discussions.
We also benefitted from interacting with Claude and ChatGPT, especially for checking numerical factors.
HT is supported by the Shoucheng Zhang Graduate Fellowship.
\appendix

\section{Faddeev-Popov measure}
\label{app:FP}
In this paper we work in the static diagonal gauge which is used in Monte Carlo \cite{Hanada:2007ti,Berkowitz:2016jlq},
\begin{equation}
\label{eq:appFP_gauge}
A(t) \;=\; \frac{1}{\beta}\,\mathrm{diag}(\alpha_1,\dots,\alpha_N),
\quad -\pi < \alpha_i \le \pi.
\end{equation}
The Faddeev-Popov measure for the $\alpha_i$ variables is
\begin{equation}
\label{eq:appFP_main}
{\;\Delta_{\rm FP}(\alpha)= \prod_{i<j}\sin^2\!\left(\frac{\alpha_i-\alpha_j}{2}\right),
\quad
S_{\rm FP} = -\,2\sum_{i<j}\log\!\left|\sin\!\frac{\alpha_i-\alpha_j}{2}\right|.\;}
\end{equation}
This is the standard form used in lattice simulations
\cite{Berkowitz:2016jlq}; we review the derivation below starting from the Faddeev-Popov procedure. %
Including FP ghosts $c,\bar c$ that are adjoint-valued and have no zero mode
(only $n\ne 0$ Matsubara modes), the static-gauge ghost action is
\cite{Kawahara:2007ib}
\begin{equation}
S_{\rm gh} =N\!\int_0^\beta\!\d t\,\tr\big(\partial_t\bar c\,D_t c\big),
\quad D_t c = \partial_t c - \ii[A_0,c],
\end{equation}
with $A_0 = \mathrm{diag}(\alpha_1,\dots,\alpha_N)/\beta$.  
To evaluate the Faddeev-Popov determinant, we
decompose $c$
into the eigenbasis of ${A_0}$ as well as into Fourier modes labelled by $n$. The off-diagonal generator $E_{ij}$
($i\ne j$) carries eigenvalue $(\alpha_i-\alpha_j)/\beta$, while the Cartan
directions carry eigenvalue zero. For each $i,j,n$ we obtain the factor
\begin{equation}
M^{(n)}_{ij} = \omega_n\!\left(\omega_n - \frac{\alpha_i-\alpha_j}{\beta}\right),
\quad \omega_n=\frac{2\pi n}{\beta}.
\end{equation}
Pairing $(i,j)$ with $(j,i)$ and $n$ with $-n$, the determinant gives
\begin{equation}
\prod_{n\ne 0}M^{(n)}_{ij}M^{(n)}_{ji}
= \prod_{n>0}\frac{(2\pi n)^4}{\beta^8}\big[(2\pi n)^2 - (\alpha_i-\alpha_j)^2\big]^2.
\end{equation}
Using the identity $\sin x = x\prod_{n\ge 1}\big(1 - x^2/(\pi n)^2\big)$,
\begin{equation}
\prod_{n>0}\!\left[(2\pi n)^2 - (\alpha_i-\alpha_j)^2\right]
\propto \frac{\sin\!\big((\alpha_i-\alpha_j)/2\big)}{(\alpha_i-\alpha_j)/2},
\end{equation}
up to an $\alpha$-independent infinite product.  The Cartan directions
contribute only an $\alpha$-independent constant.  Combining and absorbing
the residual $1/(\alpha_i-\alpha_j)^2$ factors into the Vandermonde Jacobian
that comes from diagonalizing $A_0$ onto the Cartan, the net
$\alpha$-dependent piece of the gauge-fixing measure is
\eqref{eq:appFP_main}.

In dimensionless variables $Y_\mu = (\beta/\lambda)^{1/4}(A_0, X_{i,0})$
the bosonic zero-mode action \eqref{eq:bIKKT} is independent of $\beta$ and
$\lambda$, so the holonomy
phases are
\begin{equation}
\alpha_i = \beta\,(A_0)_{i} = \lambda^{1/4}\beta^{3/4}\,(Y_0)_{i}.
\end{equation}
So in the high temperature limit, %
\begin{equation}
\Delta_{\rm FP}(\alpha) = \prod_{i<j}\sin^2\!\left(\frac{\alpha_i-\alpha_j}{2}\right)
\approx  \prod_{i<j}\left(\frac{\alpha_i-\alpha_j}{2}\right)^2
\propto \prod_{i<j}\big[(Y_0)_{i} - (Y_0)_{j}\big]^2,
\end{equation}
which is the standard flat Vandermonde for diagonalizing the Hermitian matrix
$Y_0$.  %
Combined with the bosonic zero-mode action
\eqref{eq:bIKKT}, the zero-mode measure becomes O(10)-invariant in the high temperature limit.

\section{\texorpdfstring{Large $D$ expansion}{Large D expansion}}
\label{largeDexpansion}
In this section we compute $\lr{(\mathcal{T}_{10})^2}$ in \eqref{eq:T10def} to the next order in the large-$D$ expansion, following the strategy outlined in section~\ref{sec: main text large D}.

Following \cite{Hotta:1998en}, we first review the ingredients of the large-$D$ analysis of the bosonic Yang--Mills matrix model that are needed for our computation. The model effectively behaves as a multi-flavor vector model, so one can perform a standard $1/D$ expansion. For our purposes it is convenient to rescale
\begin{equation}
Z_\mu\equiv (2D)^{1/4}Y_\mu
\end{equation}
and expand in a basis $\{T_a,\ a=1,\ldots,N^2-1\}$ of traceless Hermitian generators,
\begin{equation}
Z_\mu\equiv Z_\mu^aT^a,\ \ \ \Tr(T^aT^b)=\delta^{ab},\ \ \ T^{a}_{ij}T^{a}_{kl}=\delta_{il}\delta_{jk}-N^{-1}\delta_{ij}\delta_{kl},
\label{eq: B2}
\end{equation}
The quartic action \eqref{eq:bIKKT} can be decoupled by introducing a collective field $\Theta^{ab}$ and performing a Hubbard--Stratonovich transformation, rendering the action quadratic in $Y_\mu^a$. For fixed $\Theta^{ab}$ configuration, the matrix integral is Gaussian with two-point function
\begin{equation}
\lr{Z^{a}_\mu Z^{b}_\nu}=\delta_{\mu\nu}N^{-1}\left[\delta^{ab}+\sqrt{\frac{2}{D}}\,\Theta^{ab}+\frac{2}{D}\Theta^{ac}\Theta^{cb}+O(D^{-3/2})\right].
\label{eq: ZZ in terms of Theta}
\end{equation}
Integrating out $Z_\mu^a$ gives an effective theory for $\Theta^{ab}$. The leading large-$D$ contribution is $O(D^0)$, i.e.
\begin{equation}
\lr{Z_\mu^a Z_\nu^b}_0=N^{-1}\delta_{\mu\nu}\delta^{ab}
\end{equation}
which reproduces \eqref{YYprop}. The next-to-leading term arises at $O(D^{-1})$ and corresponds to loop corrections in the collective-field description. The Gaussian correlator of $\Theta^{ab}$ is~\cite{Hotta:1998en}
\begin{equation}
\lr{\Theta^{ab}\Theta^{cd}}\equiv V^{abcd},
\end{equation}
\begin{equation}
V^{abcd}\equiv \frac{1}{N^2-1}\left[-\frac{2}{3}NF^{abcd}+NG^{abcd}+\frac{1}{2}(\delta^{ac}\delta^{bd}+\delta^{ad}\delta^{bc})+\frac{1}{6}\delta^{ab}\delta^{cd}\right],
\label{eq: Vabcd}
\end{equation}
\begin{equation}
F^{abcd}\equiv\frac{1}{4}(f^{abcd}+f^{bacd}+f^{abdc}+f^{badc}),\ G^{abcd}\equiv\frac{1}{2}(f^{acbd}+f^{adbc}),\ f^{abcd}:=\Tr(T^aT^bT^cT^d).
\end{equation}
Although $V^{abcd}$ looks cumbersome, it is just a fixed tensor determined by $N$. The effective vertex for $\Theta^{ab}$ is given by the following action:
\begin{equation}
\begin{aligned}
\exp[-S_\text{eff,int}]
&=\exp\!\left[
\sum_{n=3}^{+\infty}
\frac{1}{n}\left(\frac{2}{D}\right)^{(n-2)/2}
\Tr(\Theta^n)
\right] \\
&=1+\frac{1}{3}\sqrt{\frac{2}{D}}\,\Tr(\Theta^3)
+\frac{1}{4}\frac{2}{D}\Tr(\Theta^4)
+\frac{1}{18}\frac{2}{D}\Tr(\Theta^3)^2
+O(D^{-3/2})
\end{aligned}
\label{eq: effective HS vertex}
\end{equation}

With these ingredients, we evaluate $\lr{(\mathcal T_{10})^2}$. Using the global $O(D)$ symmetry and \eqref{eq:T10def},
\begin{equation}
\lr{(\mathcal T_{10})^2}=\frac{9!}{(2D)^{5}}\sum_{\sigma\in S_9}\operatorname{sgn}(\sigma)\lr{\Tr(Z_0Z_1\cdots Z_9)\Tr(Z_0Z_{\sigma(1)}\cdots Z_{\sigma(9)})}
\label{eq: T10 square def}
\end{equation}
As explained in section~\ref{sec: main text large D}, the tree-level contribution comes from a single summand $\sigma=\rho$ in \eqref{eq: T10 square def}, which is the only permutation yielding a planar diagram:
\begin{equation}
\lr{(\mathcal T_{10})^2}_0=\frac{9!}{(2D)^{5}}
\end{equation}

At order $O(D^{-1})$ we must retain the $\sqrt{2/D}\,\Theta^{ab}$ and $\frac{2}{D}\Theta^{ac}\Theta^{cb}$ terms in \eqref{eq: ZZ in terms of Theta}; and the $\frac{1}{3}\sqrt{2/D}\,\Tr(\Theta^3)$ term in \eqref{eq: effective HS vertex}. (The other two $O(D^{-1})$ terms in \eqref{eq: effective HS vertex} do not contribute because they can only generate vacuum diagrams at this order.) Now, we discuss these three contributions separately.

Consider the propagator renormalization from $\frac{2}{D}\Theta^{ac}\Theta^{cb}$ and $\frac{1}{3}\sqrt{2/D}\,\operatorname{Tr}(\Theta^3)$.
For $\frac{2}{D}\Theta^{ac}\Theta^{cb}$, since this term is already $O(D^{-1})$, we may replace it by its expectation value. Using \eqref{eq: Vabcd}, we obtain
$\lr{\Theta^{ac}\Theta^{cb}}=V^{accb}=\frac{7N^2-1}{6N^2-6}\delta^{ab}$.
This amounts to a renormalization of the bare propagator and contributes
\begin{equation}
\frac{\lr{(\mathcal T_{10})^2}}{\lr{(\mathcal T_{10})^2}_0}\supset\frac{2}{D}\frac{7N^2-1}{6N^2-6}\times 10\approx\frac{70}{3D}
\label{eq: B10}
\end{equation}
For $\frac{1}{3}\sqrt{2/D}\,\Tr(\Theta^3)$, the only $O(D^{-1})$ contribution arises from contracting it with $\sqrt{2/D}\,\Theta^{ab}$ (its square contributes only to vacuum diagrams). This again renormalizes the bare propagator, since $\lr{\Theta^{ab}\Tr(\Theta^3)}=-\frac14\frac{7N^2-1}{N^2-1}\delta^{ab}$. Therefore, it contributes
\begin{equation}
\frac{\lr{(\mathcal T_{10})^2}}{\lr{(\mathcal T_{10})^2}_0}\supset-\frac{2}{D}\frac12\frac{7N^2-1}{6N^2-6}\times 10\approx-\frac{35}{3D}
\label{eq: B11}
\end{equation}

We also have to consider the vertex insertion from $\sqrt{2/D}\,\Theta^{ab}$.
Because of the $\delta_{\mu\nu}$ in \eqref{eq: ZZ in terms of Theta}, each pair $Z_\mu\otimes Z_\mu$ (which appears exactly once in \eqref{eq: T10 square def}) can supply one factor of $\sqrt{2/D}\,\Theta^{ab}$. Thus, at $O(D^{-1})$ we only need to select two distinct indices $\mu\neq\nu$ and consider the induced four-point vertex. Concretely (no sum over $\mu,\nu$),
\begin{equation}
\begin{aligned}
\lr{Y_\mu\otimes Y_\mu \otimes Y_\nu\otimes Y_\nu}
&=(T^a\otimes T^b\otimes T^c\otimes T^d)\lr{\lr{Y_\mu^aY_\mu^b}\lr{Y_\nu^cY_\nu^d}}\\
&\supset (T^a\otimes T^b\otimes T^c\otimes T^d) N^{-2}\frac{2}{D}\lr{\Theta^{ab}\Theta^{cd}}\\
&=N^{-2}\frac{2}{D}(T^a\otimes T^b\otimes T^c\otimes T^d)V^{abcd}\\
&=\frac{2N^2}{D(N^2-1)}\Big\langle Y_\mu\otimes Y_\mu \otimes Y_\nu\otimes Y_\nu\\
&\qquad\qquad\times\Big[-\frac16\underbrace{N\Tr(Y_\mu Y_\mu Y_\nu Y_\nu)}_{\textstyle \mV_1}
+\frac14\underbrace{N\Tr(Y_\mu Y_\nu Y_\mu Y_\nu)}_{\textstyle \mV_2}\Big]\Big\rangle_0\\
&+\frac{2}{D(N^2-1)}\Big\langle
\frac12Y_\mu\otimes Y_\nu \otimes Y_\mu\otimes Y_\nu
+\frac12Y_\mu\otimes Y_\nu \otimes Y_\nu\otimes Y_\mu\\
&\qquad\qquad\qquad\qquad
+\frac16Y_\mu\otimes Y_\mu \otimes Y_\nu\otimes Y_\nu\Big\rangle_0\\
\end{aligned}
\end{equation}
The three terms in the last line are suppressed by $N^{-2}$ and therefore do not contribute to planar diagrams. The planar vertex is thus given by the third line, i.e.
$\mV_1$ and $\mV_2$, arising from the $F^{abcd}$ and $G^{abcd}$ terms in \eqref{eq: Vabcd}, respectively.
Each of $\mV_1$ and $\mV_2$ contains several Wick contractions. For later diagram enumeration we display them explicitly: $\mV_1$ consists of four contractions,
\begin{equation}
\vcenter{\hbox{\includegraphics[width=0.65\linewidth]{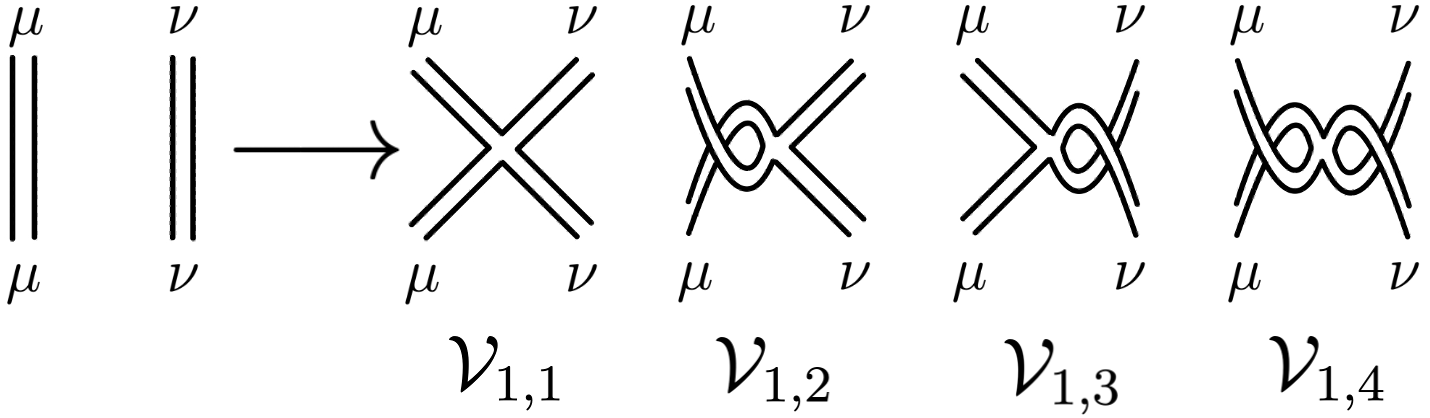}}}
\end{equation}
while $\mV_2$ consists of two contractions, each with an additional combinatorial factor of $2$,
\begin{equation}
\vcenter{\hbox{\includegraphics[width=0.4\linewidth]{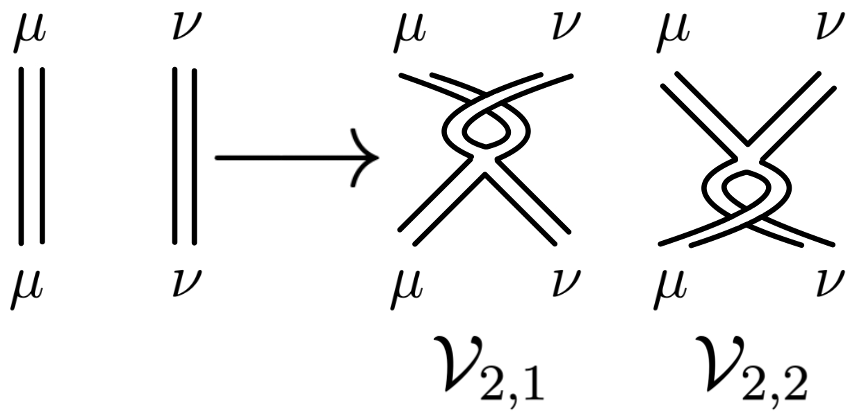}}}
\end{equation}

We now evaluate the sum in \eqref{eq: T10 square def} by separating two cases: (i) $\sigma=\rho$, where the vertex insertion must preserve planarity; and (ii) $\sigma\neq\rho$, where a vertex insertion may restore planarity.

For case (i), the only planar insertion is $\mV_{1,1}$ between nearest neighbors,
\begin{equation}
\vcenter{\hbox{\includegraphics[width=0.23\linewidth]{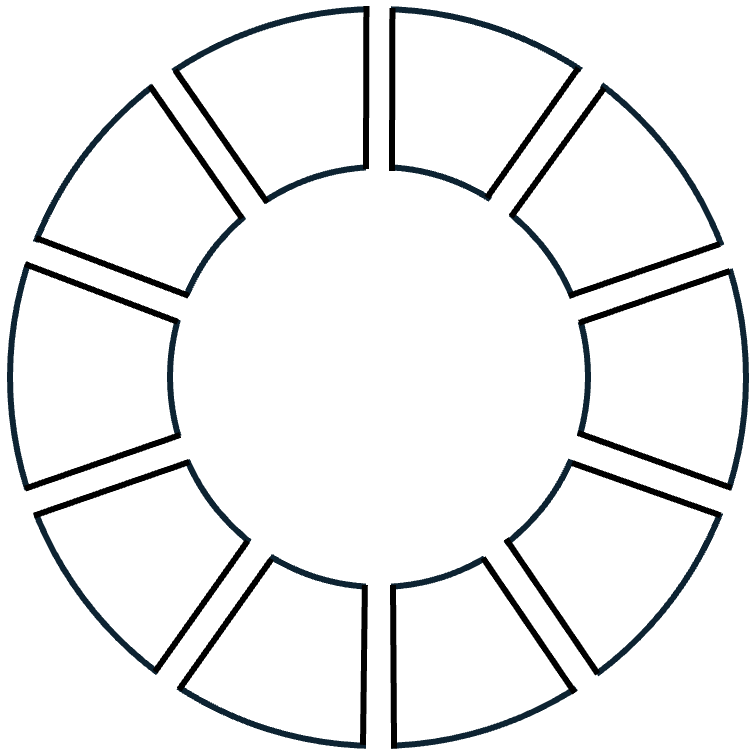}}}
\;\mathrel{\vcenter{\hbox{\scalebox{2}{$\xlongrightarrow{\mV_{1,1}}$}}}}\;
\vcenter{\hbox{\includegraphics[width=0.23\linewidth]{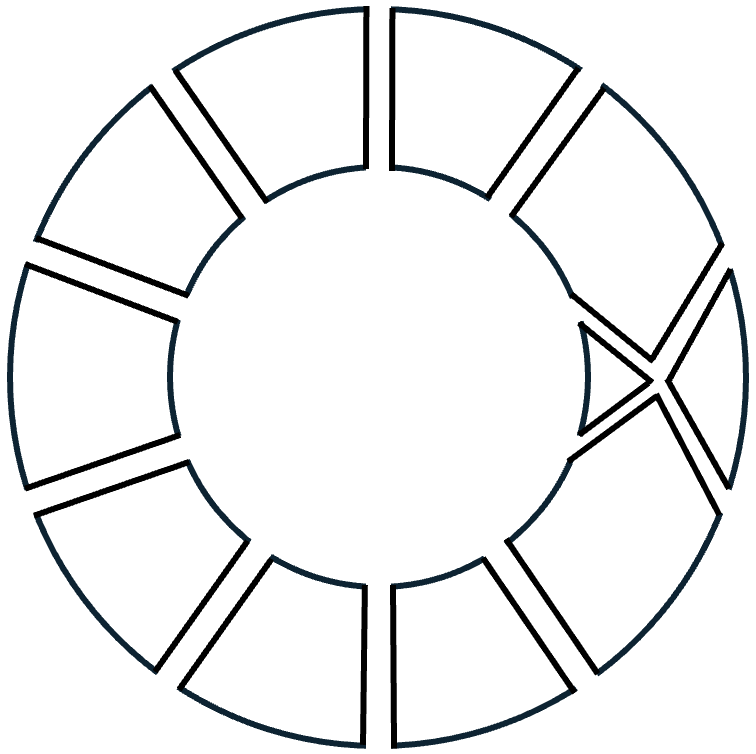}}}
\end{equation}
which contributes
\begin{equation}
\frac{\lr{(\mathcal T_{10})^2}}{\lr{(\mathcal T_{10})^2}_0}\supset\frac{2N^2}{D(N^2-1)}\left(-\frac16\right)\times10\approx-\frac{10}{3D}
\label{eq: B15}
\end{equation}
where the factor $\times10$ counts the ten nearest-neighbor pairs.

For case (ii), there are two classes of permutations $\sigma$ for which planarity can be restored. The first class consists of permutations differing from $\rho$ by a single nearest-neighbor transposition; inserting $\mV_{2,2}$ rescues planarity,
\begin{equation}
\vcenter{\hbox{\includegraphics[width=0.23\linewidth]{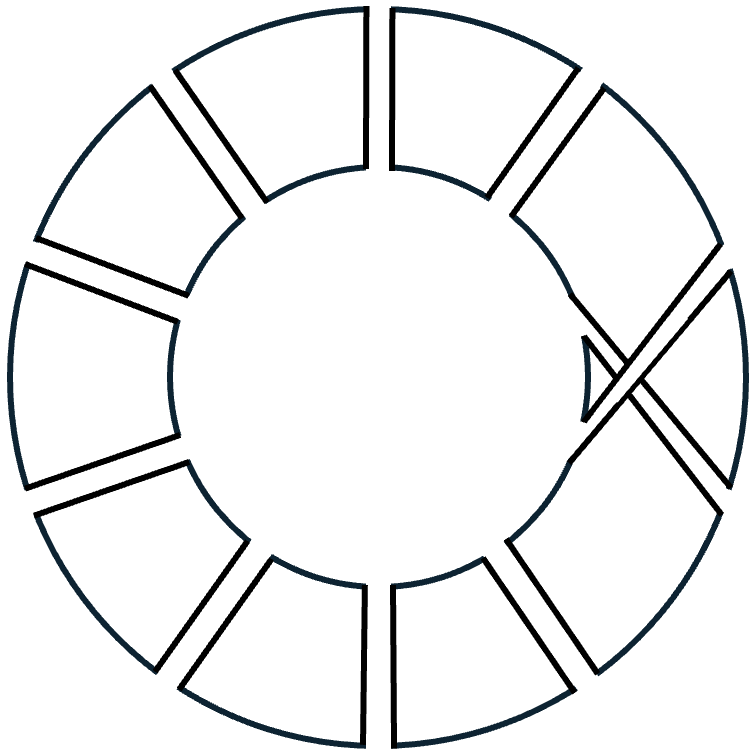}}}
\;\mathrel{\vcenter{\hbox{\scalebox{2}{$\xlongrightarrow{\mV_{2,2}}$}}}}\;
\vcenter{\hbox{\includegraphics[width=0.23\linewidth]{MC_plot/large_D_4.png}}}
\end{equation}
with contribution
\begin{equation}
\frac{\lr{(\mathcal T_{10})^2}}{\lr{(\mathcal T_{10})^2}_0}\supset\frac{2N^2}{D(N^2-1)}\left(\frac14\right)\times2\times 8\times(-1)\approx-\frac{8}{D}
\label{eq: B17}
\end{equation}
Here $\times2$ counts the two Wick contractions in $\mV_{2,2}$, $\times8$ counts the number of such permutations, and the factor $(-1)$ is $\operatorname{sgn}(\sigma)$ for a single transposition.

The second class consists of the two particular permutations
$\sigma_1=(8,7,6,5,4,3,2,1,9)$ and $\sigma_2=(1,9,8,7,6,5,4,3,2)$.
They are again rescued by inserting $\mV_{2,2}$ (illustrated for $\sigma_1$),
\begin{equation}
\vcenter{\hbox{\includegraphics[width=0.26\linewidth]{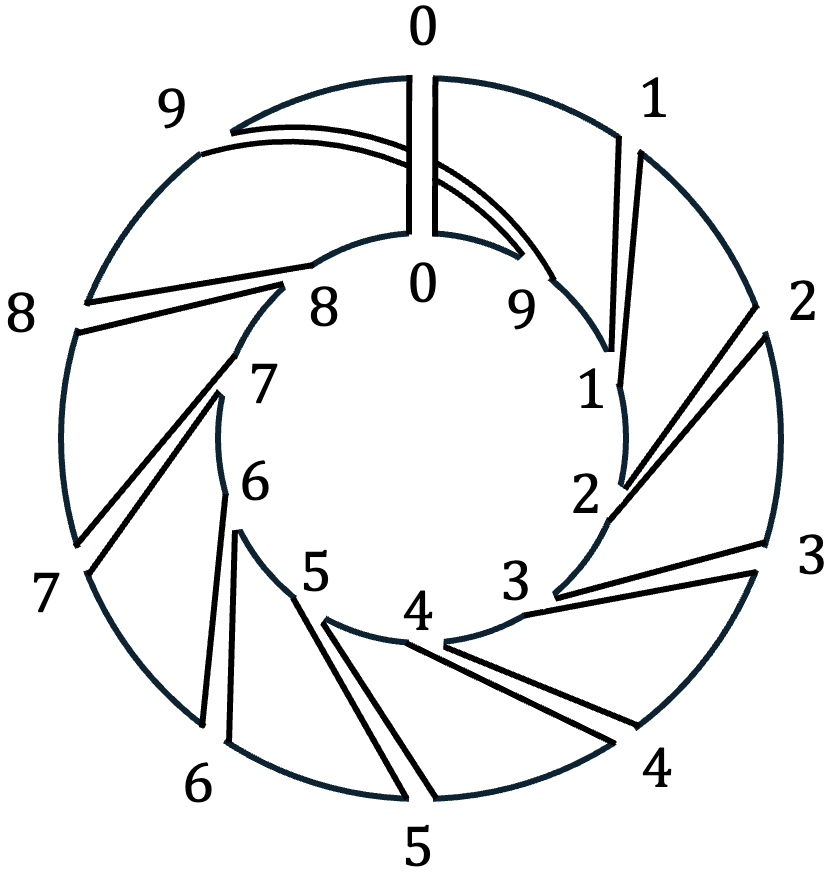}}}
\;\mathrel{\vcenter{\hbox{\scalebox{2}{$\xlongrightarrow{\mV_{2,2}}$}}}}\;
\vcenter{\hbox{\includegraphics[width=0.26\linewidth]{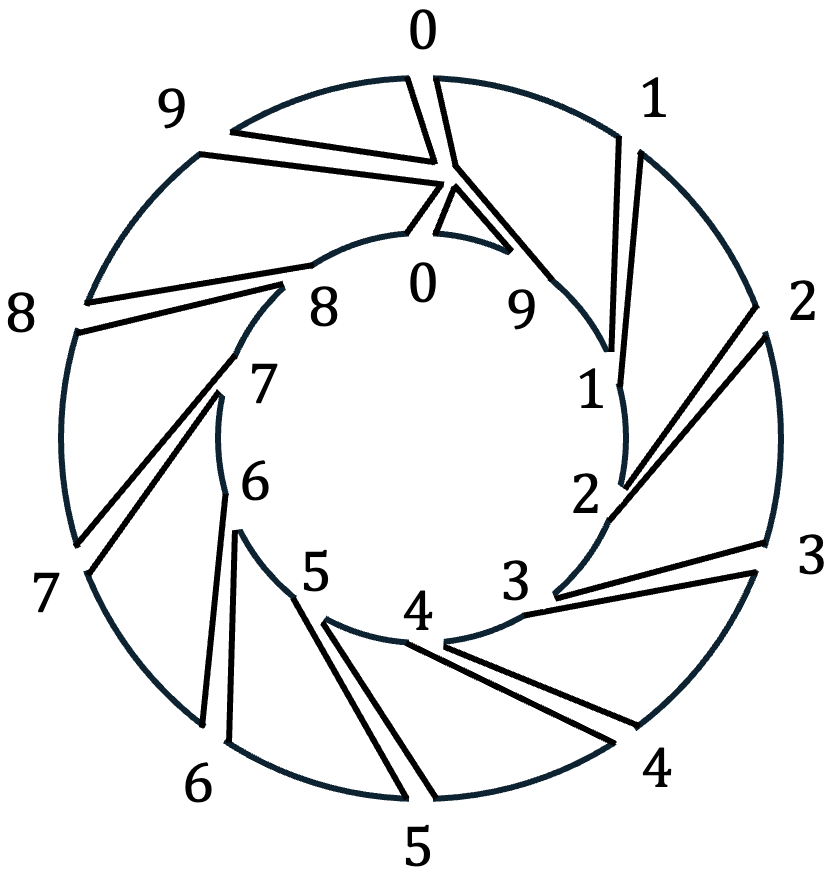}}}
\end{equation}
In this figure we have labeled the $\mu$ indices explicitly. The outer loop is read clockwise, while the inner loop is read counterclockwise. The contribution is
\begin{equation}
\frac{\lr{(\mathcal T_{10})^2}}{\lr{(\mathcal T_{10})^2}_0}\supset\frac{2N^2}{D(N^2-1)}\left(\frac14\right)\times2\times 2\times(+1)\approx\frac{2}{D}
\label{eq: B19}
\end{equation}
where $(+1)$ reflects $\operatorname{sgn}(\sigma_1)=\operatorname{sgn}(\sigma_2)=1$.

Combining \eqref{eq: B15}, \eqref{eq: B17}, and \eqref{eq: B19}, the net contribution from the $\sqrt{2/D}\,\Theta^{ab}$ term in \eqref{eq: ZZ in terms of Theta} is
\begin{equation}
\frac{\lr{(\mathcal T_{10})^2}}{\lr{(\mathcal T_{10})^2}_0}\supset-\frac{28}{3D}
\label{eq: B20}
\end{equation}

Finally, combining \eqref{eq: B10}, \eqref{eq: B11} and \eqref{eq: B20}, we obtain the full $O(D^{-1})$ correction,
\begin{equation}
\lr{(\mathcal T_{10})^2}=\lr{(\mathcal T_{10})^2}_0\left[1+\frac{7}{3D}+O(D^{-2})\right]
\end{equation}

\subsection{A code to enumerate diagrams}

To cross-check the coefficient $-\frac{28}{3D}$ in \eqref{eq: B20}, we implemented a semi-analytic enumeration of planar contributions, ensuring that no planar diagrams are missed. The code evaluates the finite-$N$ expression for $\ev{(\mathcal{T}_{10})^2}$.

We start from the contribution of the $\sqrt{2/D}\,\Theta^{ab}$ terms,
\begin{equation}
\begin{aligned}
\frac{\lr{(\mathcal T_{10})^2}}{\lr{(\mathcal T_{10})^2}_0}
&\supset \frac{2}{D}N^{-10}\sum_{\sigma\in S_9}\operatorname{sgn}(\sigma)
\sum_{0\leq r<s\leq 9}
\sum_{\{a_\mu,b_\mu\}}
\Tr(T^{a_0}\cdots T^{a_9})
\Tr(T^{b_0}T^{b_{\sigma(1)}}\cdots T^{b_{\sigma(9)}})\\
&\ \ \ \ \ \ \ \ \ \ \ \ \ \ \ \ \ \ \ \ \ \ \ \ \ \ \ \ \ \ \ \ \ \ \ \ \ \ \ \ \ \ \ \ \ \times
V^{a_r b_r a_s b_s}
\prod_{\mu\neq r,s}\delta^{a_\mu b_\mu}
\end{aligned}
\label{eq:B21}
\end{equation}
Each generator index $a_i$ and $b_j$ appears exactly twice. We can therefore iteratively apply the completeness relation (the third identity in \eqref{eq: B2}), equivalently
\begin{equation}
\begin{aligned}
&\Tr(AT^aBT^a)
=\Tr(A)\Tr(B)-\frac1N\Tr(AB),\\
&\Tr(AT^a)\Tr(BT^a)
=\Tr(AB)-\frac1N\Tr(A)\Tr(B), \ \ \ \ \ \ \forall A,B
\end{aligned}
\end{equation}
to eliminate pairs of generators.
Substituting $V^{abcd}$ into \eqref{eq:B21} and reducing traces step by step, one eventually arrives at a rational function of $N$, using $\Tr(1)=N$ at the end.

Our Python implementation gives
\begin{equation}
\begin{aligned}
\eqref{eq:B21}&=\frac{2}{D}
\frac{
-28N^{12}+890N^{10}-8344N^8+12610N^6
+161272N^4-627200N^2+460800
}{
6(N^2-1)N^{10}
}\\
&=-\frac{2}{D}\frac{(N^2-16)(N^2-9)(N^2-4)(14N^4-25N^2-400)}
{3N^{10}}\\
&\approx-\frac{28}{3D}
\end{aligned}
\end{equation}
which agrees with \eqref{eq: B20}.

The same code also extracts the finite-$N$ result at $O(D^0)$,
\begin{equation}
\frac{\lr{(\mathcal T_{10})^2}}{\lr{(\mathcal T_{10})^2}_0}=\frac{(N^2-16)(N^2-9)(N^2-4)(N^2-1)(N^2-10)}{N^{10}}+O(D^{-1})
\end{equation}
so that the finite-$N$ expression through $O(D^{-1})$ is
\begin{equation}
\begin{aligned}
\frac{\lr{(\mathcal T_{10})^2}}{\lr{(\mathcal T_{10})^2}_0}&=\frac{(N^2-16)(N^2-9)(N^2-4)(N^2-1)(N^2-10)}{N^{10}}\\
&\ \ \ +\frac{2}{D}\frac{(N^2-16)(N^2-9)(N^2-4)(7N^4-305N^2+850)}{N^{10}}\\
&\ \ \ +O(D^{-3/2})
\end{aligned}
\label{eq: B28}
\end{equation}

Note that both the $O(D^0)$ and $O(D^{-1})$ coefficients vanish for $N=2,3,4$, consistent with the identity $\mathcal{T}_{10}\equiv0$ without any ensemble averaging for these values of $N$. This vanishing is kinematic rather than dynamical, and follows from the \href{https://en.wikipedia.org/wiki/Amitsur-Levitzki_theorem}{Amitsur--Levitzki theorem}.

Concretely, for any collection of $q$ $N\times N$ matrices $\{B_1,\ldots,B_q\}$ define the anti-symmetrized product
\begin{equation}
s_{q}(B_1,...,B_q):=\sum_{\sigma\in S_q}\operatorname{sgn}(\sigma)B_{\sigma(1)}\cdots B_{\sigma(q)}
\end{equation}
The Amitsur--Levitzki theorem states $s_{2N}(B_1,...,B_{2N})=0$, as a matrix identity. Using the recursion relation
\begin{equation}
s_{q+1}(B_1,...,B_{q+1})
=\sum_{j=1}^{q+1}(-1)^{j-1}
B_j\,s_q(B_1,...,B_{j-1},B_{j+1},...,B_{q+1}),
\end{equation}
one further finds that $s_q(B_1,...,B_{q})=0$ for all $q\geq2N$. In our case $q=9$, hence the vanishing for $N=2,3,4$. 

In the next section, we directly observe through Monte Carlo that each single instance of $\mathcal T_{10}$ is zero for these values of $N$. We also benchmark the $O(1)$ and $O(D^{-1})$ coefficients for the smallest nontrivial value of $N$ ($N=5$), and the prediction from \eqref{eq: B28} matches well with Monte Carlo; see figure \ref{fig: MC}(f).

\section{Monte Carlo of the bosonic zero mode theory}
\label{MC}
In this appendix we use Monte Carlo methods to simulate the bosonic zero-mode theory defined by the action \eqref{eq:bIKKT}, with the main goal of estimating the expectation value $\langle(\mathcal T_{10})^2\rangle$ in \eqref{eq:T10def}\footnote{The definition of \eqref{eq:T10def} does not have $O(D)$ symmetry since it explicitly only involves the first ten matrices. To reduce the fluctuation in  Monte Carlo, it turns out useful to average over (at least a subgroup of) $O(D)$, the permutation $S_D$. We are allowed to do so because the final answer does not depend on which ten matrices we choose, due to $O(D)$ symmetry. Practically, $S_D$ is still too large, so we randomly sample and average over $K$ (few hundreds) out of $\binom{D}{10}$ choices of ten matrices.}. We also study the straightforward $D$-dimensional generalization of \eqref{eq:bIKKT} in order to compare with the analytic large-$D$ results in section~\ref{sec: main text large D} and appendix~\ref{largeDexpansion}. For numerical stability\footnote{In the massless bosonic IKKT model the action has classical flat directions corresponding to mutually commuting (simultaneously diagonalizable) matrices. Although these directions are lifted by quantum fluctuations of off-diagonal modes, we observe slow drift in Monte Carlo time at finite statistics. We therefore introduce a small mass term and extrapolate to the zero-mass limit.}, we introduce a mass term and subsequently extrapolate to the zero-mass limit. The most general action we consider is
\begin{equation}
\tilde{S}_{0m}=-\frac{N}{4}\Tr([Y_\mu,Y_\nu]^2)+\alpha N\Tr(Y_\mu^2),\ \mu=0,1,...,D-1
\end{equation}
When $\alpha=0$ and $D=10$, $\tilde S_{0m}$ reduces to $S_{0m}$ in \eqref{eq:bIKKT}. This model has been studied numerically in~\cite{Jha:2021exo} using the hybrid Monte Carlo (HMC) algorithm, which we follow here. We first benchmark our HMC code by measuring simple observables such as $\lr{\Tr Y_\mu^2}$ and $\lr{\Tr Y_\mu^4}$ and find agreement with~\cite{Jha:2021exo, Li:2025tub}.

\begin{figure}[H]
    \centering
    \includegraphics[width=0.99\linewidth]{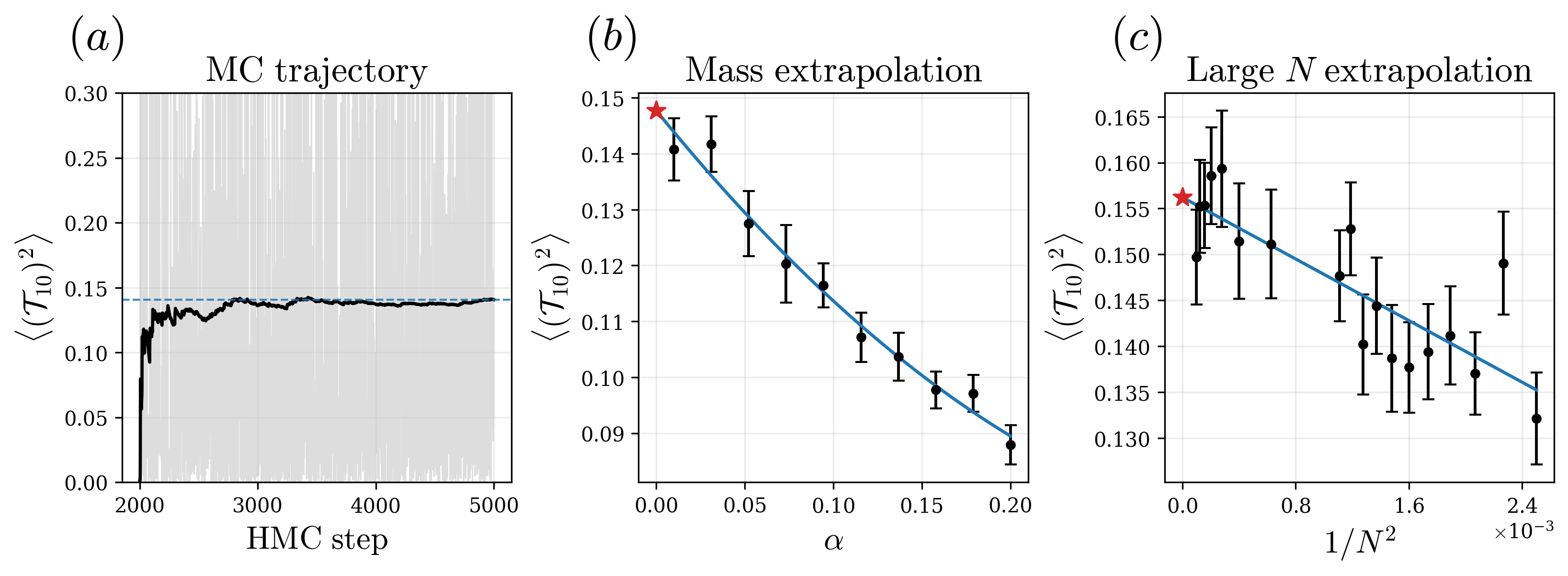}
    \includegraphics[width=0.99\linewidth]{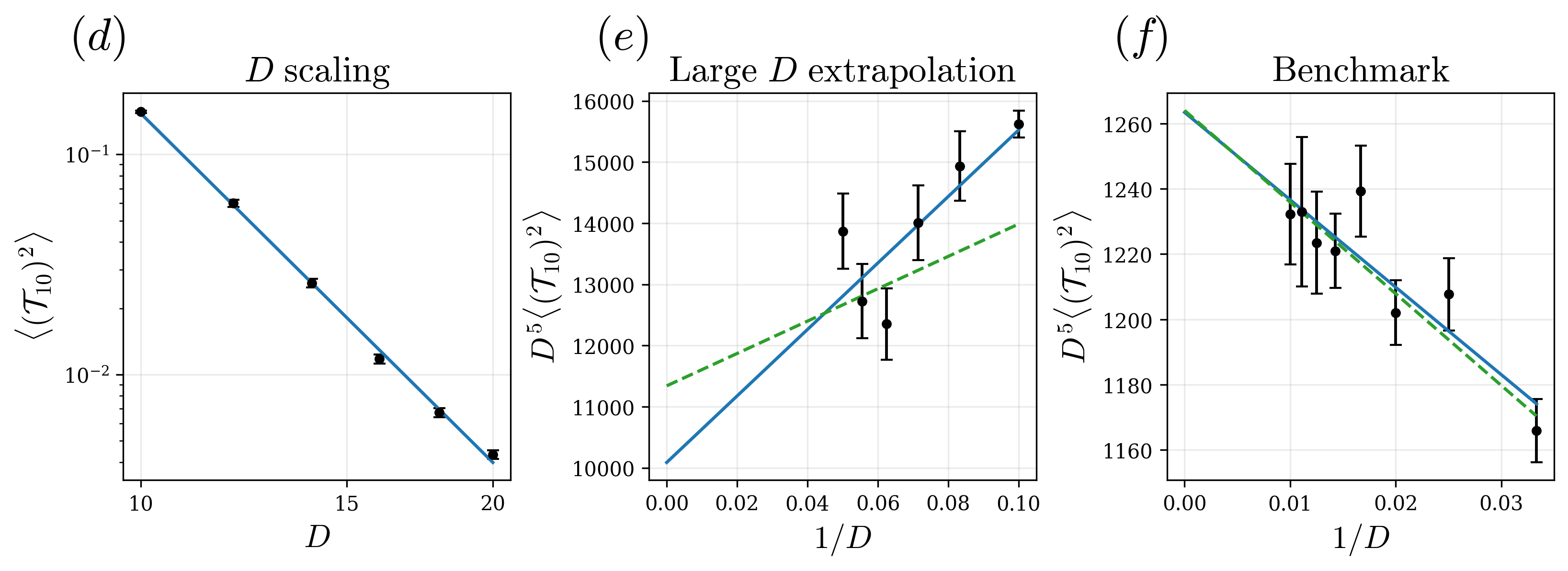}
    \caption{\textbf{(a)} Monte Carlo history at $D=10$, $N=30$, $\alpha=0.01$, showing $\lr{(\mathcal{T}_{10})^2}$ versus HMC step (after 2000 thermalization steps). The light gray curve is the instantaneous value with large fluctuation, the black curve is the stabilized running average, and the blue dashed line is the average value at the end of the HMC run. \textbf{(b)} Mass extrapolation at fixed $D=10$ and $N=30$, showing $\lr{(\mathcal{T}_{10})^2}$ versus mass $\alpha$. Black points with error bars are HMC data, the blue curve is a polynomial fit, and the red star is the extrapolated value at $\alpha=0$. \textbf{(c)} Large-$N$ extrapolation at fixed $D=10$ using data already extrapolated to $\alpha=0$, showing $\lr{(\mathcal{T}_{10})^2}$ versus $N^{-2}$. Black points with error bars are the mass-extrapolated data, the blue line is a linear fit in $N^{-2}$, and the red star is the extrapolated value at $N^{-1}=0$. \textbf{(d)} Large-$D$ scaling of $\lr{(\mathcal T_{10})^2}$ after extrapolation to $\alpha=0$ and $N^{-1}=0$, obtained from a linear fit (blue line) on a log-log plot. We find $\lr{(\mathcal T_{10})^2}\sim D^{-5.26}$. \textbf{(e)} Finer $D$-dependence compared with the analytic large-$D$ expansion: a fit (\textcolor{blue!90!black}{blue line}) of $D^5\lr{(\mathcal T_{10})^2}\approx c_1(1+c_2D^{-1})$ yields $c_1\approx10087$ and $c_2\approx5.38$, which should be compared with the large-$D$ analytic calculation (\textcolor{green!60!black}{green dashed line}) with $c_1=11340$ and $c_2=7/3$. \textbf{(f)} Small $N$ benchmark of the large-$D$ analytic calculation for $\lr{(\mathcal T_{10})^2}$ in \eqref{eq: B28}. At $N=5$, it predicts (\textcolor{green!60!black}{green dashed line}) $D^5\lr{(\mathcal T_{10})^2}\approx c_1(1+c_2D^{-1})$ with $c_1=493807104/390625\approx1264.15, c_2=-20/9\approx-2.22$, whereas from fitting (\textcolor{blue!90!black}{blue line}) the Monte Carlo data we obtain $c_1\approx1263.42, c_2\approx-2.12$.
    }
    \label{fig: MC}
\end{figure}

The model parameters are $(D,N^{-1},\alpha)$, and we wish to estimate $\lr{(\mathcal T_{10})^2}$ at $(10,0,0)$. For each fixed $D$ we generate data at several values of $N$ and $\alpha$ (figure~\ref{fig: MC}(a)). We first extrapolate to $\alpha=0$ (figure~\ref{fig: MC}(b)) and then to $N^{-1}=0$ (figure~\ref{fig: MC}(c)). With the unnormalized trace convention in \eqref{eq:T10def}, $\lr{(\mathcal T_{10})^2}\sim O(1)$ in the large-$N$ limit, while subleading non-planar corrections admit an expansion in powers of $N^{-2}$.

We begin with $D=10$. After extrapolating in $\alpha$ and $N$, we obtain $\lr{(\mathcal{T}_{10})^2}\approx0.156$ ({\color{red} red star} in figure~\ref{fig: MC}(c)). We observe that the large-$D$ calculation to $O(D^{-1})$ in appendix~\ref{largeDexpansion} predicts $\lr{(\mathcal{T}_{10})^2}\approx\frac{9!}{20^5}\times(1+\frac{7}{30})\approx0.140$, in reasonable agreement.

We next study the dependence on $D$. From the log-log fit in figure~\ref{fig: MC}(d) we find $\lr{(\mathcal{T}_{10})^2}\sim D^{-5.26}$, close to the large-$D$ prediction $\sim D^{-5}$. To further compare the coefficient, in figure~\ref{fig: MC}(e) we fit \begin{align}
    \lr{(\mathcal{T}_{10})^2}\approx c_1 D^{-5} (1+c_2D^{-1}), \quad c_1\approx 1.01 \times 10^{4}, \quad c_2\approx5.38.
\end{align}  The corresponding large-$D$ prediction is $c_1=\frac{9!}{2^5}=11340$ and $c_2=\frac73\approx2.33$, again in reasonable agreement. {For the error bars in all panels of figure~\ref{fig: MC}, we use the estimate $\sqrt{\tau_\text{auto}\sigma^2/n_\text{step}}$, where $n_\text{step}$ is the total number of HMC steps, $\tau_\text{auto}$ is the autocorrelation time, and $\sigma^2$ is the sample variance. This choice is motivated by viewing $n_\text{step}/\tau_\text{auto}$ as the effective number of independent samples and then applying standard error propagation. We estimate $\tau_\text{auto}$ from the  autocorrelation function $f_l:=c_3\sum_{t=1}^{n_\text{step}-l}\delta x_t\delta x_{t+l}$ (the normalization $c_3$ is chosen such that $f_0=1$) via the windowed sum $\tau_\text{auto}\approx\sum_{l=1}^{w}f_l$, where $w$ is a cutoff. Here $\delta x_t$ denotes the deviation of the data from its mean. This is because, for an exponentially decaying correlation $f_l\sim e^{-l/\tau_\text{auto}}$, one has $\sum_lf_l\sim\int dl\cdot e^{-l/\tau_\text{auto}}\sim \tau_\text{auto}$.}

\section{Higher dimensions}\label{app:higherD}

\begin{figure}[H]
\centering
\begin{tikzpicture}[scale=1.2,baseline={(0,0)}]
    \def\Rout{3.6}\def\Rin{3.4}\def\rout{1.2}\def\rin{1.0}
    \def\angOut{1.35}\def\angIn{3.85}
    \def\bone{1.9}\def\btwo{2.7}
    \def\angBtwo{1.9}\def\angBone{2.7}

    \draw[blue,line width=1.2pt] (0,0) circle (\Rout);
    \foreach \a in {0,36,72,108,144,180,216,252,288,324} {
        \draw[blue,line width=1.2pt] ({90+\a+\angOut}:\Rin) arc ({90+\a+\angOut}:{90+\a+36-\angOut}:\Rin);
        \draw[red,line width=1.2pt] ({90+\a+\angIn}:\rout) arc ({90+\a+\angIn}:{90+\a+36-\angIn}:\rout);
        \begin{scope}[rotate around={\a:(0,0)}]
            \draw[black,line width=1.1pt] (-0.08,\Rin) -- (-0.08,\btwo);
            \draw[black,line width=1.1pt] (0.08,\Rin) -- (0.08,\btwo);
            \draw[black,line width=1.1pt] (-0.08,\rout) -- (-0.08,\bone);
            \draw[black,line width=1.1pt] (0.08,\rout) -- (0.08,\bone);
        \end{scope}
    }

\foreach \i in {0,...,9} {
        \begin{scope}[rotate around={-36*\i-18:(0,0)}]
    \node[blue] at (0,\Rout+0.3) {$\tilde{k}_\i$};
    \node[red] at (0,\rin-0.3) {$k_\i$};
    \end{scope}
    \begin{scope}[rotate around={-36*\i-10:(0,0)}]
    \node at (0,\rout+0.3) {${q}_\i$};
    \end{scope}
    \begin{scope}[rotate around={-36*\i-6:(0,0)}]
        \node at (0,\btwo+0.25) {$\tilde{q}_\i$};
    \end{scope}

}
    \draw[red,line width=1.2pt] (0,0) circle (\rin);
            \fill[even odd rule,pattern={Lines[angle=45,distance=5pt,line width=1pt]},pattern color=gray] (0,0) circle (\btwo) (0,0) circle (\bone);
    \foreach \a in {0,36,72,108,144,180,216,252,288,324} {
        \draw[black,line width=1.1pt]
            ({90+\a+\angBtwo}:\btwo)
            arc ({90+\a+\angBtwo}:{90+\a+36-\angBtwo}:\btwo);
        \draw[black,line width=1.1pt]
            ({90+\a+\angBone}:\bone)
            arc ({90+\a+\angBone}:{90+\a+36-\angBone}:\bone);
    }
\end{tikzpicture}
\caption{The higher-dimensional ring. We have labeled the $d-1$ spatial momentum carried by each propagator. The bosonic modes that enter the hashed annulus carry momentum $q$ or $\tilde{q}$ but are zero-modes on the thermal circle.}
\label{fig:higherD}
\end{figure}
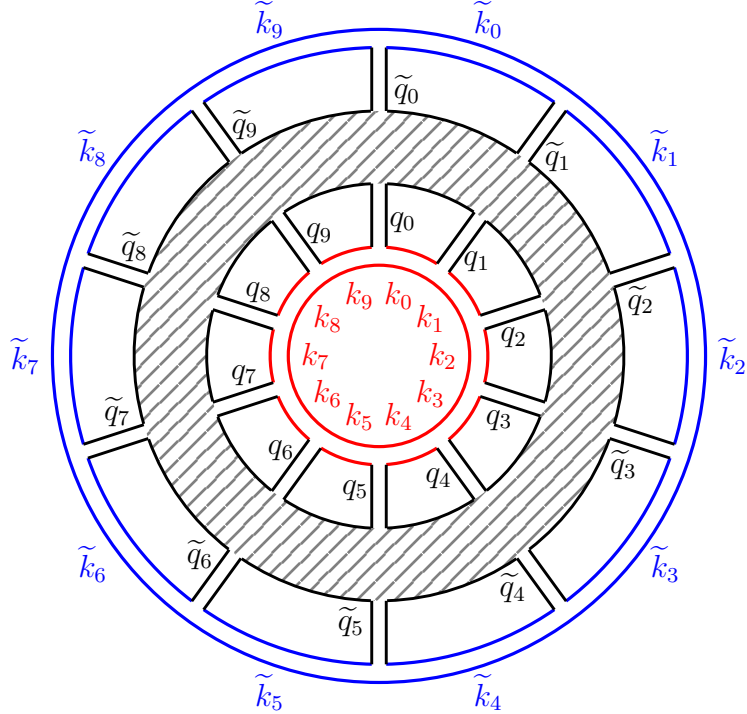

Let us consider maximally supersymmetric Yang-Mills in $d$ dimensions for $d=2,3,4$ on $M_d = \mathbb{R}^{d-1}\times S^1_\beta$.
We denote worldvolume indices as $ 
\mu,\nu = 0,1,\ldots,d-1$, the spatial parts of the worldvolume indices as $ 
a,b = 1,\ldots,d-1$ 
and scalar indices as $I,J = 1,\ldots,10-d$. The zero-th direction is the Euclidean thermal circle. Bosons are periodic
and fermions are antiperiodic around this circle.
The Euclidean action is: %
\begin{align}
\label{eq:SYM_d}
S_d
&=
\frac{N}{\lambda}
\int_0^\beta \dd\tau
\int \dd^{d-1}x\,
\tr\Bigg\{
    \frac14 F_{\mu\nu}F_{\mu\nu}
    +\frac12 (D_\mu X_I)(D_\mu X_I)
    -\frac14 [X_I,X_J]^2
\nonumber\\
&\hspace{7em}
    +\frac12 \Psi^T
    \left(
        D_0-\ii\gamma^aD_a-\gamma^I[X_I,\cdot]
    \right)
    \Psi
\Bigg\}.
\end{align}
Here $D_\mu=\partial_\mu-\ii[A_\mu,\cdot\,]$. $\gamma^a$, $\gamma^I$ are the $9$ real symmetric $16\times 16$
gamma matrices satisfying $\{\gamma^i,\gamma^j\} = 2\delta^{ij}$, with
$i,j$ running over both spatial and scalar directions. 
We can again consider the replica trick, now for this higher dimensional setup. This gives similar diagrams as before, see Figure \ref{fig:higherD}.
In this diagram, the spatial momenta satisfy
\begin{align}
    k_i &= k_{i-1} - q_i,\\
    \tilde{k}_i &= \tilde{k}_{i-1} - \tilde{q}_{i-1}.
\end{align}
Just as for BFSS, we work in static gauge, where $A_\tau$ is independent of time. In the above diagram, the insertion associated with $q_0$ or $\tilde{q}_0$ is the $A_\tau$ insertion, while all other vertices are associated with an
$X_i$ or a gauge field $A_x$ coming from the covariant derivative %
\begin{align} \label{cderiv}
    \slashed{D}(q_i) = \slashed{q}_i - \ii \,  \ad \, \slashed{A}(q_i). 
\end{align}
In addition to the spatial momentum, each fermionic propagator is associated with a Matsubara frequency $\omega_r$ for the red ring and $\omega_{r}'$ for the blue ring. We consider only Matsubara zero modes for the bosonic degrees of freedom (in black); hence, time-translation invariance enforces that all Matsubara frequencies for propagators that are part of a given ring are equal. 
\subsection{Position space computation}
While performing the integrals exactly in momentum space seems daunting, in position space the 
small-$\beta$ position-space expansion is simple.  The fermion Kaluza-Klein mass is
of order $1/\beta$, so each propagator is localized on distances of order
$\beta$.  With the normalization of \eqref{eq:SYM_d}, the reduced propagator for
mode $r\in\mathbb Z+\frac12$ is
\begin{equation}
  G_r(x-y)
  =
  \frac{1}{2\pi\ii r}\,
  \delta^{(d-1)}(x-y)\,\mathbf 1_{16}
  +
  \frac{\ii\beta}{(2\pi\ii r)^2}\,
  \gamma^a\partial_{x^a}\delta^{(d-1)}(x-y)
  +O(\beta^2\partial^2\delta).
  \label{eq:G_delta_expansion_d}
\end{equation}
The leading delta function is the identity in spinor space; the second term is
the first derivative correction and carries one $\gamma^a$.

Now specialize to $d=2$.  Let us first consider a ring with 9 insertions and hence 9 fermion propagators, connecting one $A_0$ insertion and eight scalar insertions. %
At order $\beta^{10}$, one segment must either use the derivative term in
\eqref{eq:G_delta_expansion_d}, or contain an additional insertion of the
spatial gauge field $A_x$.  These two possibilities can occur on any segment
and combine into the covariant combination \eqref{cderiv}
\begin{align}
-\ii \gamma_x D_x=\gamma_x(-\ii\partial_x-\ad \, A_x).
\end{align}  After integrating by parts, this gives a covariant derivative acting on one field in the local ring.

Now, both the derivative correction and the $A_x$ insertion carry
$\gamma^x$.  Moving this $\gamma^x$ to the beginning of the spinor trace gives a
factor $(-1)^m$ after crossing $m$ scalar gamma matrices.  Therefore the
Leibniz sum for the derivative on the $j$th scalar is
$\sum_{m=0}^{j-1}(-1)^m$, which is nonzero only for odd $j$.  Choosing the
representative with no derivative on $A_0$, the finite-$N$ operator is
\begin{equation}
  {\cal O}_{10}^{(2d)}(x)
  =
  \epsilon_{I_1\cdots I_8}
  \sum_{\ell=0}^{3}
  \tr_{\rm ad}\!\left[
    (\ad \, A_0)
    (\ad \, X_{I_1})\cdots
    (\ad \, (-\ii D_xX_{I_{2\ell+1}}))\cdots
    (\ad \, X_{I_8})
  \right].
  \label{eq:O10-2d}
\end{equation}
With the orientation convention
$\tr_\gamma(\gamma^x\gamma^{I_1}\cdots\gamma^{I_8})
=16\,\epsilon_{I_1\cdots I_8}$, the coefficient is fixed by the same
Matsubara sum used above.  The derivative term contributes one corrected
propagator $(2\pi\ii r)^{-2}$ and eight leading propagators, while the
$A_x$ insertion splits one segment into two leading propagators; as explained below \eqref{cderiv}, all the Matsubara frequencies are equal on a given ring, yielding a prefactor\footnote{Note that the second term in \eqref{eq:G_delta_expansion_d} comes with a factor of $1/(2\pi i r)^2$. However, we obtain such a term starting from 9 propagators. So we get $1/(2\pi i r)^{10}$. On the other hand, an $\ad \, A_x$ insertion involves 10 vertices and 10 propagators, also giving the same factor of $1/(2\pi i r)^{10}$. }
$(2\pi\ii r)^{-10}$:
\begin{equation}
  \theta_{10}^{(2d)}
  =
  8s_{10}\,\beta^{10}\int\dd x\,{\cal O}_{10}^{(2d)}(x),
  \qquad
  s_{10}=\sum_{r\in\mathbb Z+\frac12}\frac{1}{(2\pi\ii r)^{10}}
  =-\frac{31}{1451520}.
  \label{eq:theta10_2d_position}
\end{equation}

We have focused on the $d=2$ case for concreteness, but one can similarly work out the $d=3, 4$ cases. This result is essentially consistent with \eqref{theta2D}, except by using the position space propagators, there is no ambiguous singular term. For future work, it would be interesting to compute $\ev{\theta^2}$, again by either Monte Carlo of the lower $d-1$ bosonic theory or by a large $D$ expansion.

\subsection{Symmetry analysis}

We can also perform a symmetry analysis of the Pfaffian and see that the operator $O_{10}$ is the leading operator consistent with three symmetries.

In $d=2$, we consider the Pfaffian $\Pf(M)$ of the matrix
\begin{align}
  {\cal M}
  =
  D_0-\ii\gamma^xD_x-\gamma^I\ad \, X_I,
  \qquad
  D_\mu=\partial_\mu-\ii \, \ad \, A_\mu,
  \qquad I=1,\ldots,8.
\end{align}
We will show that 
\begin{align}
    \Pf(\mathcal{M}(X_g,A_g))=\Pf(\mathcal{M}(X,A))^*,
\end{align}
where $g$ is a group element of three different $\mathbb{Z}_2$ symmetries, $g \in \mathbb{Z}_2 \times \mathbb{Z}_2 \times \mathbb{Z}_2$.

First, one can show using similar considerations to Section \ref{sec:reflection} that under a spatial (worldvolume) reflection
    \begin{align}
    &\mathcal{R}: A_a(x) \mapsto -A_a(-x), \quad %
    \mathcal{R}: X_I(x) \to -X_I(-x), \quad 
    \mathcal{R}: A_\tau(x) \to A_\tau(-x),
\end{align}
the Pfaffian goes to its complex conjugate. 

Second, under time reversal (in the $d$-dimensional theory)
$\mathcal{T}[A_\tau(x)] = - A_\tau(x)$ the Pfaffian also goes to its complex conjugate.

Third, under a target space reflection, the Pfaffian goes to its complex conjugate. Together with SO(8) invariance, this implies that the phase is an O(8)
pseudoscalar.  To see this, fix one scalar direction \(K\), and define the transformation %
\begin{equation}
  {\cal R}_K:%
  X_J\mapsto -X_J\quad (J\neq K),
\end{equation}
leaving all other $X$ and $A$ fields fixed.
This flips seven scalar directions, and hence is an improper \(O(8)\) rotation.
Now in the fermionic path integral,
\begin{equation}
  \operatorname{Pf}{\cal M}[{\cal R}_K[X]]
  =
  \int D\Psi\,
  \exp\!\left[
    -\frac12\int \dd\tau\,\dd x\,
    \Psi^T{\cal M}[{\cal R}_K[X]]\Psi
  \right],
\end{equation}
we can change variables \(\Psi= S \Psi'= \gamma^x\gamma^K\Psi'\).  The Jacobian is
trivial, and we also have %
\begin{equation}
  S^T\gamma^xS=-\gamma^x,\quad
  S^T\gamma^KS=-\gamma^K,\quad
  S^T\gamma^JS=\gamma^J\quad (J\neq K),
  \quad S=\gamma^x\gamma^K .
\end{equation}
Therefore
\begin{equation}
  S^T{\cal M}[{\cal R}_K[X]]S
  =
  D_0+\ii\gamma^xD_x+\gamma^I\ad \, X_I
  =
  {\cal M}[X]^*,
\end{equation}
where in the last step we used the fact that the gamma matrices are real and that
\((\ad \, Y)^*=-\ad \, Y\), so \(D_\mu^*=D_\mu\).  

In addition to these symmetries, the Pfaffian must be invariant under Poincar\'e transformations and should be a gauge-invariant operator. Combining this with our EFT expectation that the Pfaffian should be an integral of a local operator, one can argue that the leading operator consistent with these properties should contain one $A_0$ insertion (to be $\mathcal{T}$-odd), one $D_x$ insertion (to be $\mathcal{R}$-odd) and 8 $X$ insertions to contract with the $\epsilon$ symbol, in order to be an SO(8) pseudoscalar.

This argument does not fix the relative coefficients of the terms in the sum \eqref{eq:O10-2d}; it would be interesting to understand if there are further symmetries that enforce these terms.

\section{Subleading temperature correction}
\label{sec:subleading_T}
\newcommand{\calO}{\mathcal O}
We now estimate the relative size of the next correction to
$\log\langle\cos\theta\rangle$ in the high-temperature expansion. Two
distinct sources contribute at the same parametric order; each
arises with one extra factor of $\varepsilon^2=(\lambda\beta^3)^{1/2}$
relative to the leading $\varepsilon^{20}$ piece. 

\subsection{Higher order operators}
The first source of higher order terms comes from expanding the log of the Pfaffian to higher orders in $\beta$, while still truncating to bosonic zero modes of $Y$. 
The discrete symmetries that enforce $\theta$ to be the
$\varepsilon^{10}$ pseudoscalar $\cten\,\mathcal O_{10}$ permit two more
operators which are functions of the zero modes at $O(\varepsilon^{12})$:
\begin{equation}
  \theta[Y]
  =
  \cten\,\varepsilon^{10}\calO_{10}[Y]
  +
  b_{12}\,\varepsilon^{12}
  \left(\calO_{3,9}[Y]-\calO_{1,11}[Y]\right)
  +O(\varepsilon^{14}),
  \qquad
  b_{12}=\frac{16}{24}s_{12}
  =
  \frac{691}{479001600}.
\end{equation}
Here \(s_{12}=\sum_{r\in\mathbb Z+\frac12}(2\pi\ii r)^{-12}\) is the Matsubara sum from the 12 fermionic propagators in the expansion of $\Tr\log\mathcal M$. We use the notation
\begin{equation}
  T_{\nu_1,\nu_2,\cdots,\nu_{12}}
  =
  \Tr_{\rm ad}\!\left[
    (\ad \, Y_{\nu_1})(\ad \, Y_{\nu_2})\cdots(\ad \, Y_{\nu_{12}})
  \right].
\end{equation}
The \((1,11)\) operator is

\begin{align}
  \calO_{1,11}
  &=
  12\epsilon_{i_1\cdots i_9}
  \sum_{a=1}^{9}
  \sum_{0\le r\le s\le 9}
  (-1)^{s-r} T_{0,i_1\cdots i_r,a,i_{r+1}\cdots i_s,a,i_{s+1}\cdots i_9} 
  .
\end{align}
Here there are two distinguished insertions with repeated indices $Y_a$ that are placed in all possible locations in the word.\footnote{Using the fact that the trace of eleven $\gamma$ matrices is non-zero only if every one of the nine $SO(9)$ indices appears an odd number of times, we can show that the two extra $X_i$'s that are inserted in $\sim A_0 (X_i)^{11}$ must have the same index.} 
Similarly, the \((3,9)\) operator is
\begin{align}
  \calO_{3,9}
  =
  \epsilon_{i_1\cdots i_9}
  \sum_{0\le r\le s\le t\le 9}
  T_{i_1\cdots i_r,0,i_{r+1}\cdots i_s,0,i_{s+1}\cdots i_t,0,i_{t+1}\cdots i_9}.
\end{align}
This is the sum over the three \(Y_0\) insertion locations.

\begin{figure}[t]
\centering
\begin{tikzpicture}[scale=.8,baseline={(0,0)}]
    \def\Rout{3.6}\def\Rin{3.4}\def\rout{1.2}\def\rin{1.0}
    \def\angOut{1.35}
    \def\angIn{3.85}
    \draw[blue,line width=1.2pt] (0,0) circle (\Rout);
    \foreach \a in {0,36,72,108,144,180,216,252,288,324} {
        \draw[blue,line width=1.2pt] ({90+\a+\angOut}:\Rin) arc ({90+\a+\angOut}:{90+\a+36-\angOut}:\Rin);
        \draw[red,line width=1.2pt] ({90+\a+\angIn}:\rout) arc ({90+\a+\angIn}:{90+\a+36-\angIn}:\rout);
    }
    \foreach \a in {72,108,144,180,216,252,288,324} {
        \begin{scope}[rotate around={\a:(0,0)}]
            \draw[black,line width=1.1pt] (-0.08,\Rin) -- (-0.08,\rout);
            \draw[black,line width=1.1pt] (0.08,\Rin) -- (0.08,\rout);
            \draw[black,line width=1.1pt,fill=black] (0,{0.5*(\Rin+\rout)}) circle (0.15);
        \end{scope}
    }
    \foreach \a/\lab/\pos/\xshift in {0/{$n\!\neq\!0$}/right/0pt,36/{$-n$}/above right/-5pt} {
        \begin{scope}[rotate around={\a:(0,0)}]
            \draw[black,line width=1.1pt,decorate,
                  decoration={snake,segment length=5pt,amplitude=1.6pt}]
                (-0.10,\Rin) -- (-0.10,\rout);
            \draw[black,line width=1.1pt,decorate,
                  decoration={snake,segment length=5pt,amplitude=1.6pt}]
                (0.10,\Rin) -- (0.10,\rout);
            \draw[black,line width=1.1pt,fill=black] (0,{0.5*(\Rin+\rout)}) circle (0.15);
            \node[\pos=3pt,xshift=\xshift] at (0,{0.5*(\Rin+\rout)}) {\lab};
        \end{scope}
    }
    \draw[red,line width=1.2pt] (0,0) circle (\rin);
\end{tikzpicture}
\caption{Same diagram as figure~\ref{fig:replica_nonperturbative_internal}, except
that two adjacent bosonic rungs carry non-zero Matsubara frequencies $n$ and $-n$. We have replaced two $X_i$ matrices with fast modes; the gauge field $A_0$ is constant in static gauge.
}
\label{fig:replica_nonzero_mode}
\end{figure}
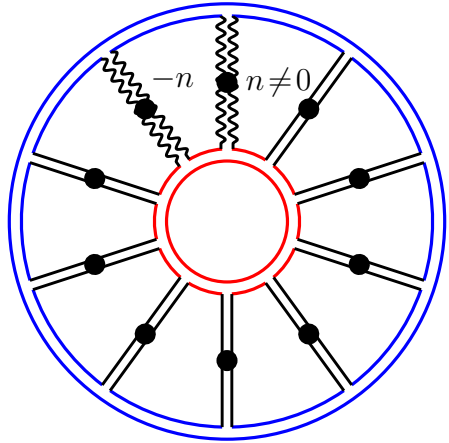

One might also worry about corrections to the phase which come from
Matsubara non-zero modes of the bosonic matrices. Replacing one zero-mode rung with a ``fast mode''
does not contribute, since energy conservation cannot be satisfied. %
The leading diagram with fast modes is therefore one with two rungs replaced by fast modes, depicted in
figure~\ref{fig:replica_nonzero_mode}, where one bosonic mode has frequency
$n \ne 0 $ and the other has frequency $-n$. This gives an extra factor of
\(\varepsilon^4\) relative to the leading \(O(\varepsilon^{20})\) term.
Thus non-zero spatial modes first contribute to
\(\log\langle\cos\theta\rangle\) at
\(O(\varepsilon^{24})=O((\lambda\beta^3)^6)\), not at
\(O(\varepsilon^{22})=O((\lambda\beta^3)^{11/2})\).

\subsection{Order-\texorpdfstring{\(\varepsilon^2\)}{eps2} correction to the measure}
At leading order, the bosonic zero-mode measure is given by $\prod \d Y_i  e^{-S_0(Y)}$ where $S_0$ is the action of BFSS truncated to the bosonic zero modes. More generally, we can write the measure as $\prod \d Y_i e^{-S_\text{eff}(Y)}$, with the $O(\varepsilon^2)$ correction to the effective action
\begin{equation}
  S = S_0(Y) + \varepsilon^2 S_{\rm eff}^{(2)}(Y), \quad 
 S_{\rm eff}^{(2)} =
  -\frac43 N\left\{\sum_{i=1}^9\tr(Y_iY_i)-\tr(Y_0Y_0)\right\}.
  \label{eq:Seff2}
\end{equation}
This is the next-to-leading high-temperature insertion of
eq.~(4.12) in ref.~\cite{Kawahara:2007ib}, translated to our conventions.\footnote{ The authors \cite{Kawahara:2007ib} write the insertion as
\[
  \calO_{\rm KNT}
  =
  -\left(\frac{d-1}{12}-\frac p8\right)
  N\beta^{3/2}
  \left\{\tr(\widetilde A_i)^2-\tr(\widetilde A_D)^2\right\}.
\] 
For BFSS,
\(d=9\) and \(p=16\), so the coefficient is \(4/3\).  Finally
\(\widetilde A_D=Y_0\), \(\widetilde A_i=Y_i\), and
\(V_2=2N\{\sum_i\tr(Y_iY_i)-\tr(Y_0Y_0)\}\), hence
\(\calO_{\rm KNT}=(2/3)\varepsilon^2V_2\), where we have restored the 't Hooft coupling. }
This operator is generated by accounting for the interactions of the bosonic and fermionic non-zero modes with the zero modes. After integrating out the bosonic and fermionic non-zero modes with a Gaussian measure, the result gives new terms in the effective action for the zero modes.

The operator in \cite{Kawahara:2007ib} which corrects the effective action was derived for the thermal BFSS theory without the phase-quenched approximation. However, the effective action at this order does not differ between phase-quenched and unquenched.

The operator \(S_{\rm eff}^{(2)}\) is even under both \(Y_0\to -Y_0\) and
spatial orientation reversal, so it cannot generate a one-point function
for \(\calO_{10}\); it first affects the phase average through
\(\langle\calO_{10}^2\rangle\).

\subsection{Combined results}

Now let us combine the correction from the phase operator and the correction
to the effective action. Keeping only terms through order
\((\lambda\beta^3)^{11/2}\), the cumulant expansion gives
\begin{align}
  \log\langle\cos\theta\rangle
  &=
  -\frac{\cten^2}{2}(\lambda\beta^3)^5
  \langle\calO_{10}^2\rangle_{0,c}
  \nonumber\\
  &\quad
  +
  (\lambda\beta^3)^{11/2}
  \left[
    \frac{\cten^2}{2}
    \langle\calO_{10}^2S_{\rm eff}^{(2)}\rangle_{0,c}
    -
    \cten b_{12}\,
    \langle\calO_{10}\calO_{3,9}\rangle_{0,c}
    +
    \cten b_{12}\,
    \langle\calO_{10}\calO_{1,11}\rangle_{0,c}
  \right]
  +O\!\left((\lambda\beta^3)^6\right).
  \label{eq:subleading-final}
\end{align}
The numerical coefficients appearing in \eqref{eq:subleading-final} are
\begin{equation}
  \frac{\cten^2}{2}
  =
  \frac{961}{65840947200},
  \qquad
  \cten b_{12}
  =
  \frac{21421}{86910050304000}.
\end{equation}
We see that the corrections are expected to be a power series in
\((\lambda\beta^3)^{1/2}\) multiplying the leading
\((\lambda\beta^3)^5\) term.
To proceed, one would have to estimate the three new correlators using either Monte Carlo or a large $D$ expansion. 
At large $N$, the resulting connected correlators could again be simplified by using large $N$ factorization. We leave this to future work. It would be interesting to see if the subleading correction is positive or negative, as this would give some indication of whether the sign problem becomes more or less severe at strong coupling.

\bibliography{main}
\bibliographystyle{JHEP}
\end{document}